\documentclass[draft,jgrga ]{AGUTeX}

 \usepackage[dvips]{graphicx}
\authorrunninghead{CORRAL and TURIEL}

\titlerunninghead{Variability of N. Atlantic hurricanes}

\authoraddr{
\'Alvaro Corral,
Centre de Recerca Matem\`atica,
Edifici Cc, Campus Bellaterra,
E-08193 Barcelona,
Spain.
}

\authoraddr{
Antonio Turiel,
Institut de Ci\`encies del Mar, CSIC, 
P. Mar\'{\i}tim Barceloneta 37, E-08003 Barcelona,
Spain.
}

\begin{document}

%% ------------------------------------------------------------------------ %%
%
%  TITLE
%
%% ------------------------------------------------------------------------ %%

\title{
%%Variability of individual-hurricane PDI
Variability of North Atlantic hurricanes:
seasonal versus individual-event features
}

%
% e.g., \title{Terrestrial ring current:
% Origin, formation, and decay $\alpha\beta\Gamma\Delta$}
% You may use \\ to break the title over several lines.

%% ------------------------------------------------------------------------ %%
%
%  AUTHORS AND AFFILIATIONS
%
%% ------------------------------------------------------------------------ %%

%Use \author{\altaffilmark{}} and \altaffiltext{}

% \altaffilmark will produce footnote;
% matching altaffiltext will appear at bottom of page.
% May use \\ to start a new line.

%\authors{R. C. Bales, \altaffilmark{1}
%E. Mosley-Thompson, \altaffilmark{2} R. Williams, \altaffilmark{3}
%and J. R. McConnell\altaffilmark{4}}

%\altaffiltext{1}{Department of Hydrology and Water Resources,
%University of Arizona, Tucson, Arizona, USA.}

%\altaffiltext{2}{Department of Geography, Ohio State University,
%Columbus, Ohio, USA.}

%\altaffiltext{3}{Department of Space Sciences, University of
%Michigan, Ann Arbor, Michigan, USA.}

%\altaffiltext{4}{Division of Hydrologic Sciences, Desert Research
%Institute, Reno, Nevada, USA.}

%% ------------------------------------------------------------------------ %%
%
%  ABSTRACT
%
%% ------------------------------------------------------------------------ %%

% >> Do NOT include any \begin...\end commands within
% >> the body of the abstract.

\begin{abstract}
Tropical cyclones are affected by a large number of climatic factors,
which translates into complex patterns of occurrence.
The variability of annual metrics of tropical-cyclone activity 
has been intensively studied, in particular since the sudden
activation of the North Atlantic in the mid 1990's.
We provide first a swift overview on previous work by diverse authors about
these annual metrics for the North-Atlantic basin,
where the natural variability of the phenomenon, the existence of trends,
the drawbacks of the records, and the influence of global warming 
have been the subject of interesting debates.

Next, we present an alternative approach that does not focus
on seasonal features but on the characteristics of 
single events [Corral et al., {\it Nature Phys.} 6, 693 (2010)]. 
It is argued that the individual-storm
power dissipation index (PDI) constitutes a natural way to
describe each event, and further, that the PDI statistics yields
a robust law for the occurrence of tropical cyclones
in terms of a power law.
In this context, methods of fitting these distributions are discussed.

As an important extension to this work we introduce a distribution 
function that models the whole range of the PDI density
(excluding incompleteness effects at the smallest values),
the gamma distribution, consisting in a power-law with an exponential
decay at the tail.
The characteristic scale of this decay, represented by the cutoff parameter, 
provides very valuable information on the finiteness size of the basin,
via the largest values of the PDIs that the basin can sustain.
We use the gamma fit to evaluate the influence of sea surface temperature (SST)
on the occurrence of extreme PDI values, 
for which we find an increase around 50 \%
in the values of these basin-wide events for a 0.49 $^\circ$C SST average difference.

Similar findings are observed for the effects 
%%of the El Ni\~no Southern oscillation,
of the positive phase of 
the Atlantic multidecadal oscillation and the number of hurricanes in a season
on the PDI distribution.
In the case of the El Ni\~no Southern oscillation (ENSO),
positive and negative values of the multivariate ENSO index
do not have a significant effect on the PDI distribution;
however, when only extreme values of the index are used,
it is found that the presence of El Ni\~no 
decreases the PDI of the most extreme hurricanes.
\end{abstract}

%% ------------------------------------------------------------------------ %%
%
%  BEGIN ARTICLE
%
%% ------------------------------------------------------------------------ %%

% The body of the article must start with a \begin{article} command
%
% \end{article} must follow the references section, before the figures
%  and tables.

\begin{article}

%% ------------------------------------------------------------------------ %%
%
%  TEXT
%
%% ------------------------------------------------------------------------ %%

\section{Introduction}

% empezado 14/1/2011

%Tropical cyclones is the generic name 
%encompassing typhoons, hurricanes, 
%tropical storms, and tropical depressions.
%%for what is otherwise called, 
%%depending on geographical 

Tropical cyclones are a rare phenomenon, 
with less than 100 occurrences per year worldwide 
(tropical depressions not counted)
-- just compare with $10^4$ earthquakes with magnitude larger than $4$
per year.
%(hurricane-category storms plus tropical storms,
%tropical depressions not counted).
Despite this scarcity, the societal impact
of these atmospheric systems is huge,
as it is perceived year by year by the general public.
An important issue for planning %% and mitigation
of tropical-cyclone damage mitigation is the 
knowledge of the temporal variability
of the phenomenon,
%%%of occurrence in time,
and to try to establish which part of the variability is natural 
and which part can be due to global warming.

When counting these meteorological monsters, a bit of terminology
is useful \citep{Emanuel_book}.
Tropical cyclones is the generic name 
encompassing typhoons, hurricanes, 
tropical storms, and tropical depressions
(when the context allows it, 
we will also use the vague term ``storm'').
Typhoons and hurricanes (and severe cyclonic storms) 
are ``mature'' tropical cyclones
(the difference in name is only of geographical origin),
with winds strong enough to achieve a category from 1 to 5
in the Saffir-Simpson scale \citep{Saffir_Simpson}.
In contrast, tropical storms have not so large wind speeds, 
and for tropical depressions these are even weaker.
The thresholds separating these 3 classes
are 34 and 64 knots for the so-called
maximum sustained (1 min) surface (10 m) wind speed (1 kt $=$ 0.5144 m/s),
further thresholds define the Saffir-Simpson category.
Then, speed is a way to define intensity,
and intensity at a given instant determines
the stage of the tropical cyclone (hurricane, tropical storm, etc.)
and the category in case of hurricanes or typhoons.
The stage corresponding to the maximum lifetime intensity 
determines how the storm is finally labeled.
%%, as a tropical storm, hurricane, etc.
Generally, the databases in use only contain 
more or less complete information on
tropical storms and hurricane-like systems, 
and therefore, when talking about tropical cyclones,
tropical depressions will be excluded in this text.
Note that in some references tropical depressions
are not even considered as tropical cyclones, 
they are just formative or decaying
stages of tropical storms and hurricanes
\citep{Neumann}.

Although the North Atlantic ocean 
(including the Gulf of Mexico and the Caribbean)
comprises %% an average of 
only about 10 \% of the 
tropical cyclones in the world,
this is the ocean basin which has been more
extensively studied, with routine aircraft reconnaissance
since 1944 and an intensive scrutiny of ship log records
for previous years
(satellite imagery started to be available from the 1960's
\citep{Neumann}).
This has yield %% which has lead to 
the longest and most reliable database for %%long-term trends in 
tropical cyclones,
extending back in time, with obvious limitations,
for more than one century and a half.

In this work we analyze tropical cyclone activity 
in the North Atlantic.
First, previous work on temporal variability 
is overviewed, from the point of view of direct observations, 
and the controversy about if
a global-warming signal can be separated
from the natural fluctuations
is briefly mentioned.
Fully in-depth reviews on this topic, including modeling, theory,
and paleoclimatic studies, 
have been published by \cite{Shepherd_Knutson},
\cite{Knutson_geosci} and \cite{Knutson_landsea_emanuel}.
Second (Sec. 3), an approach based on the study of the
features of individual tropical cyclones is 
presented, based on the calculation of their power dissipation index
(PDI).
Power-law fits for the PDI distribution are presented and discussed 
also,
and a superior, more general fitting function (the gamma distribution) 
is introduced in Sec. 4, taking into account the exponential decay of the tail
of the distribution.
Sections 5 and 6 analyze the effect of tropical sea surface temperature 
and diverse climatic indices (El Ni\~no, NAO, AMO, number of tropical cyclones)
on the PDI distribution.

\section{Variability of hurricane activity}

\begin{figure*}
 \noindent\includegraphics[width=39pc]{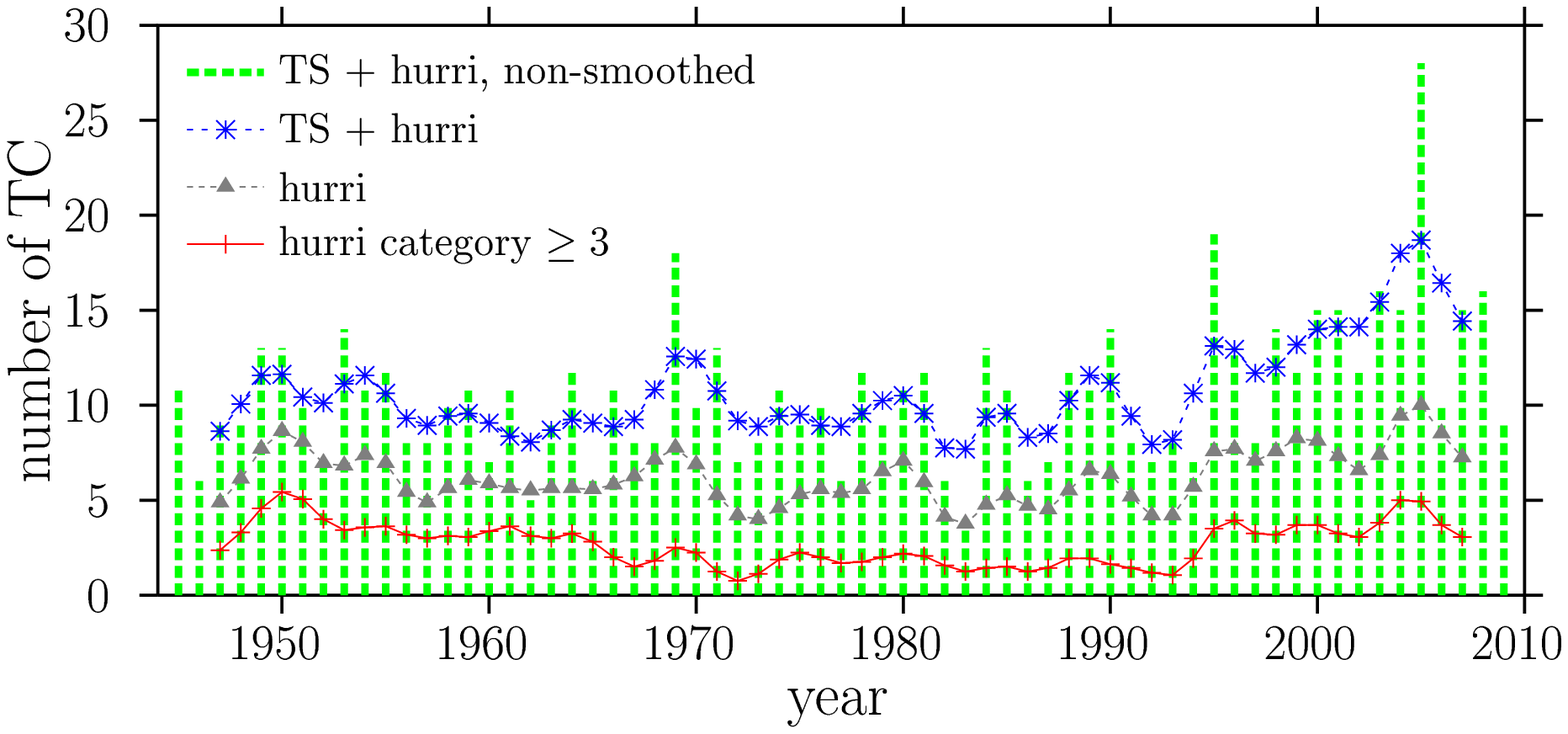}
 \caption{
Annual number of tropical cyclones, hurricanes, and major hurricanes
in the North Atlantic as a function of time.
The data has been smoothed with a 1-2-1 filter
applied twice.
Unsmoothed data are also shown for the number of tropical cyclones
(tropical storms plus hurricanes).
Corrections are not performed, neither in the counts nor in the intensities.
For the original plot see \cite{Goldenberg_Science} or \cite{Webster_Science}.
\label{fig1}}
 \end{figure*}

The large temporal variability of North-Atlantic tropical-cyclone 
frequency is clearly seen in the existing records, 
with a maximum of 28 occurrences in 2005 
(including a subtropical storm) %% or 19 in 1995 
versus only 4 in 1983, for instance,
see Fig. \ref{fig1}.
Excluding tropical storms, %%one counts
there was a maximum of
15 hurricanes also in 2005 and just 2 in 1982;
and for major hurricanes, which are those
with category 3 or above (96 kt threshold),
there were 8 in 1950
versus none in 1994
\citep{Neumann,Gray_HDP,Elsner_Kara,Bell_2005,NOAA_faq}.

But high and low tropical-cyclone activity come in multidecadal clusters.
\cite{Gray_90} reported that
the number of major hurricanes 
decreased to less than one half from the period 1947-1969 to 
1970-1987 (3.3 per year versus 1.55),
which was interpreted as a regional manifestation
of a global-scale climate variation
governed by thermoaline processes.
If an hypothetical overestimation of speeds previous to 1970 is corrected,
the decreasing trend still persists for these periods
\citep{Landsea93}; nevertheless, see also \cite{Landsea_comment}
with regard the necessity of this adjustment.

\begin{figure*}
 \noindent\includegraphics[width=39pc]{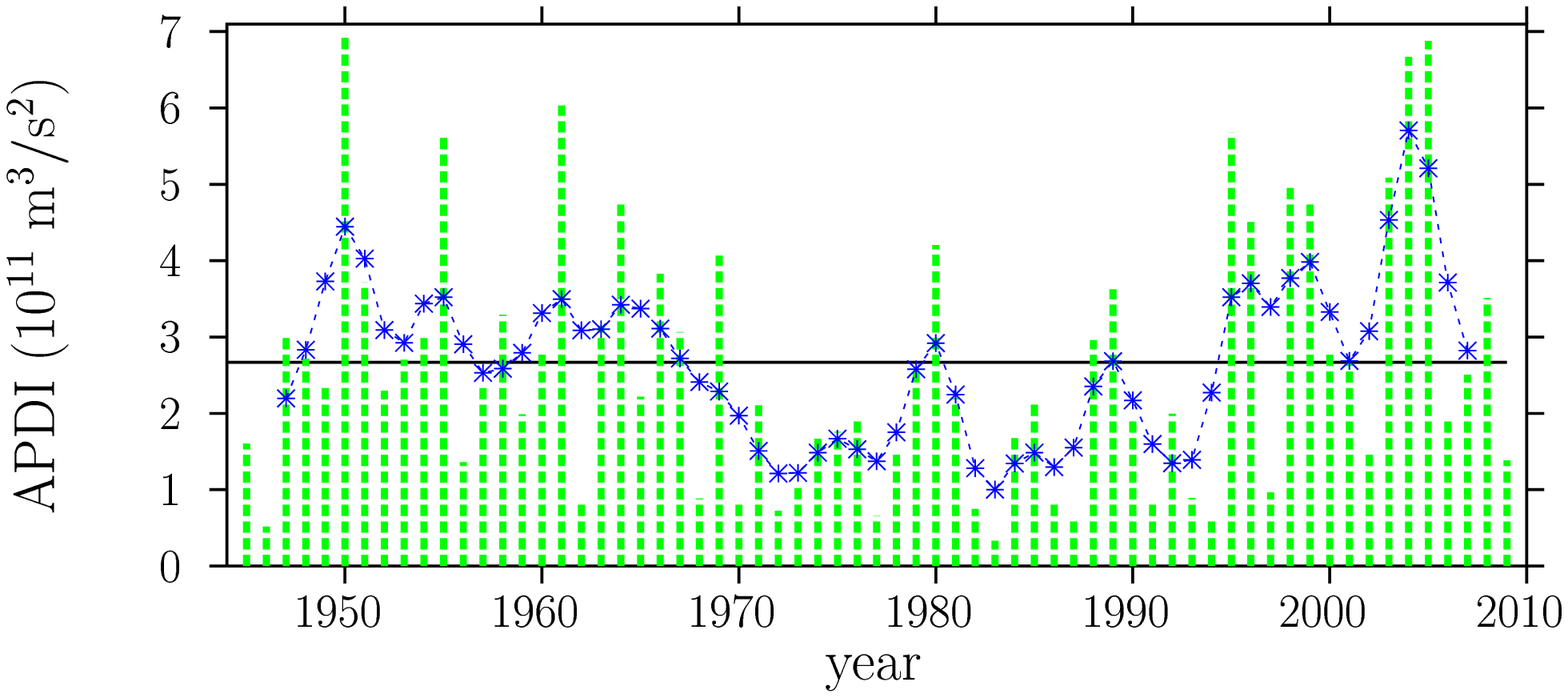}
 \caption{
Annual accumulated PDI (APDI) in the North Atlantic,
both for the original series and for the smoothed one
(by means of a 1-2-1 filter applied twice).
No corrections have been applied.
The mean of the original series is also shown.
The year with the largest APDI is 1950, but followed
very closely by 2005.
The maximum of the smoothed series is for 2004.
For the original plot see \cite{Emanuel_nature05}.
\label{fig2}}
 \end{figure*}

\cite{Elsner_Jagger00} and \cite{Goldenberg_Science}
realized that in 1995 a new period of 
high activity seemed to have started in the North Atlantic.
In the first case, the increase was related to 
the North Atlantic oscillation (NAO),
whereas in the second it was associated
to simultaneous increases in North
Atlantic sea-surface temperatures (SST) and decreases in vertical wind shear,
governed by variations of natural origin of the so-called Atlantic multidecadal mode
(or oscillation, AMO, \citep{Kerr_AMO}).
In this way, the period from 1965 to 1994 witnessed no more than 3 major
hurricanes per year, whereas in 1995 there were 5 of such storms, 
and 6 in 1996.
The return of high levels of activity was labeled as dramatic.

Later, \cite{Webster_Science} found that an increasing trend 
for the period 1970-2004 was 
also significant for hurricanes of at least category 1, in clear contrast to 
the behavior of other basins worldwide (in which tropical SST also increased).
Remarkably, 
these authors also extended the increase of major tropical cyclones to the
other basins (excluding category 3)
and noted that this was not
inconsistent with climate simulations showing
that a doubling of atmospheric CO$_2$ should lead to an increase in the
frequency of the most intense tropical cyclones.
However, subsequent analysis have claimed that the 
sign in the trend of the Northeastern Pacific 
is dependent on the time window selected
and that the increase in other basins could be an artifact 
due to the lack of quality and homogeneity of the records.
In any case, the increased activity of North-Atlantic
hurricanes of categories 4 and 5
was clear and robust \citep{Klotzbach,Kossin}.

However,
\cite{Landsea_Eos07} has shown that
the analysis of tropical-cyclone variability using
counts of events requires enormous caution.
Before the advent of aircraft reconnaissance 
in the Atlantic, in 1944, detection of open-ocean storms
was dependent on chance encounters with ships.
But even after the 1940's, aircrafts were covering essentially the west half
of the basin, so systems developing on the east part
were not always observed.
The comparison between the number of landfalling storms 
(which is assumed reliable since 1900)
and the total number of storms gives a quite stable
ratio of 59 \% since 1966 (when satellite imagery
started to operate), in contrast to a value around 75 \%
for previous years.
In addition, new operational tools, as QuikSCAT, that have became
available in the turn of this century, have allowed the identification
of one additional tropical cyclone per year (on average).
This has led Landsea to estimate a deficit in the annual frequency of storms
of 3.2 events per year up to 1965
and 1 per year from 1966 to 2002.
(One can wonder why the correction to this problem is solved by adding a constant
term, rather than multiplying by some factor.)
%%The results of this analysis appear on Fig. 1.

Obviously, storm counts do not provide a complete characterization
of tropical-cyclone activity, and other indicators are necessary.
(Here we will use the term activity in a broad sense, 
and not as a synonymous of frequency or abundance.)
In this way, total storm days, 
%defined as the sum of durations of all storms of a given type
%during a year, 
defined by \cite{Gray_HDP} as the total time all storms in a year
spend in a given storm stage, 
was found by \cite{Gray_90} to quadruple between low and high activity periods
for the stage of major hurricanes 
(a reduction to 2.1 days per year from a previous average of 8.5, 
for the periods mentioned above).
Note that this metric is different to the total duration of 
major hurricanes (which is the sum of all their durations, in any stage).
For the total duration of North Atlantic hurricanes (categories 1 to 5,
and including tropical-storm stages),
\cite{Webster_Science} reported a significant trend from 1970 to 2004.

Another quantity that has been employed 
is the net tropical cyclone activity (NTC), 
explained in the supplementary information
of \cite{Goldenberg_Science} as an average between 
storm counts and total storm days, considering
tropical storms, non-major hurricanes, and major hurricanes, 
but giving more
weight to latter and less weight to the former storms.
In this way, information on frequency, 
duration, and intensity is combined into a single number,
resulting that the NTC between 1995-2000
is twice the value of 1971-1994,
which in turn was a factor 1.6 smaller than for 1944-1970.

%%To be more concrete, the NTC counts the number of tropical storms, 

A significant step forward was taken by
\cite{Bell}, who defined the accumulated cyclone energy
(ACE) by summing, for all tropical cyclones in a season, 
the squares of the 6-hour maximum sus­tained wind speed 
for all records in which the systems were 
above the tropical storm limit (i.e., speed $\ge 34$ kt).
%%Note that 
The ACE is a more natural way to combine frequency, intensity 
and duration than the NTC, 
but it should not be interpreted as a kinetic energy
(rather, it could be the time integral of something akin to kinetic energy,
as its name denotes).
%this would be a continuous variable, 
%assuming that either the speeds or the durations were continuous.
It is worth mentioning that previously, \cite{Gray_HDP} had introduced
a related quantity, the hurricane destruction potential (HDP),
with the only difference that tropical-storm stages were not
taken into account.

Updating the results of Bell et al. 
for the ACE as an annual indicator, 
\cite{Trenberth}
noticed that between 1995 and 2004 the North Atlantic activity 
had been ``above normal'', with the only exception of the
El Ni\~no years of 1997 and 2002, and that
this increase was in parallel to the fact that that
decade was the one with the highest SST on record in the tropical 
North Atlantic (by more than 0.1 $^\circ$C)
and also to the increased water-vapor content
over the global oceans.
(The results for ACE were confirmed later by \cite{Klotzbach}.)
Although many uncertainties remain on how 
this and other environmental changes can affect
the self-organization process of tropical-cyclone formation,
once formation has taken place it seems clear that
these conditions provide more energy to the storm,
which suggests more intense winds and heavier rainfalls
under a global warming scenario.

A similar analysis was performed by \cite{Emanuel_nature05},
who introduced what he called power dissipation index (PDI),
but that we will call annual or accumulated PDI (APDI),
defined as the ACE with the main difference that
the square of the speeds is replaced by their cube
(and also, the time summation was performed for the whole 
life of the storm and not only for tropical-storm and 
hurricane stages).
It is notable that the APDI constitutes a ``proxy'' estimation
of the kinetic energy dissipated by all tropical cyclones in one season 
(not of their power).
Emanuel showed how a well-known formula for the dissipation, 
used previously by \cite{Bister_Emanuel}, could be converted into the 
APDI formula under reasonable assumptions.
Indeed, the dissipated power is given by the integral over
surface of the cube of the velocity field
(multiplied by the air density and the drag coefficient, 
which can be assumed as constants). First, accepting a similar shape 
for all storms (the same functional form for the velocity field, 
scaled by the radius of the storm and the maximum-in-space instantaneous speed)
this power is proportional to the square of the radius multiplied
by the cube of the maximum speed
(with the same proportionality constant for all storms).
Second, 
noticing the weak correlation between storm dimensions
and speed, one could assign a common radius to all storms
(in all stages) and %% one would only 
get only random errors 
in the estimation of the power. 
As the energy dissipated by a tropical cyclone is given by the time integral 
(for all its life) of its power,
replacing the integral by a discrete
summation (as the records come in discrete intervals)
one gets that the dissipated energy is roughly proportional then
to the sum of the cube of the maximum sustained wind speed,
which is (in contrast to the usage of Emanuel), what we call PDI.
Further summation for all storms provides the APDI,
for which it is expected that the errors coming from the assignment
of a common radius will more or less compensate.
See \cite{Corral_Elsner} for a non-verbal version
of this derivation, and the appendix here for an estimation
of the proportionally constant linking PDI and dissipated energy.

%In summaru, if one accumulates the PDI for all tropical cyclones
%in a season one gets an estimation of the overall
%energy dissipated in that season, as the proportionality factor is
%not relevant for comparative purposes.
%We call such metric, in contrast to Emanuel, APDI
%(accumulated or annual PDI),
%and keep the name PDI for individual dissipation
%of tropical cyclones.

In equations,
$$
APDI=\sum_{i=1}^{n} PDI_i
$$
where $i$ counts the $n$ tropical cyclones in the season
under consideration.
For a single storm, its PDI is
$$
PDI=\sum_{\forall t} v_t^3 \Delta t,
$$
where $t$ labels discrete time, $v_t$ is the maximum sustained 
surface wind speed at $t$, and we introduce a time interval
$\Delta t$ that is constant, in principle, and equal to 6 hours,
just to make the PDI independent on changes on $\Delta t$ 
%%--non-existent in the best-track records).
(nevertheless, it is an open question how a better time resolution 
can influence the stability of the PDI value, this will depend on
the smoothness of the time-evolution of the speed).
Under these assumptions, the SI units of the PDI are m$^3$/s$^2$
(for instance, hurricane Katrina [2005] yields 
PDI$=6.5 \cdot 10^{10}$ m$^3$/s$^2$).

Note that for tropical cyclone $i$,
$$
PDI_i=\langle v_t^3 \rangle_i T_i,
$$
where $\langle v_t^3 \rangle_i$
is the average of the cube of the (6-hour)
maximum sustained wind speed over storm $i$,
and $T_i$ is its duration.
Also, for year $y$,
$$
APDI_y=n_y \langle PDI \rangle_y = n_y \langle T \rangle_y \langle v_t^3 \rangle_y
$$
where $n_y$ is the number of tropical cyclones 
in that year and $\langle ... \rangle_y$
is the average of the considered quantity for the same year
(the average is performed in a different way for
the duration, of which there are $n_y$ data,
than for the speed, of which there are $\sum_i T_i/\Delta t$ records).

The previous equations show that the PDI of a tropical cyclone
is the product of its duration and its average intensity,
if one redefines intensity as the cube of the maximum sustained
surface wind speed,
whereas the APDI is the product of frequency, average duration, 
and annually averaged (6-hour) intensity.
Both the PDI and the APDI turn out to be very natural and convenient ways to estimate
tropical-cyclone activity, as they are a rough estimation of individual and
total dissipated energy, respectively.
Further, they are more robust than other measures
of storms, as duration or track lengths,
due to the fact that for these latter, the value is highly influenced
by the definition of when a tropical cyclone starts, 
whereas for the PDI and APDI such values have little influence on the 
final result, as they are weighted by the smallest values of $v_t^3$.

What Emanuel found was a clear correlation between 
``unprecedented'' increases in tropical SST and APDI,
both for the North Atlantic (more than doubled in the last 30 years) and Northwestern Pacific
(75 \% increase),
when a 1-2-1 filter was applied twice to both signals
(i.e., a 3-year running average where the central point has
a weight that is equal to that of both neighbors).
As the upswing in SST has been generally ascribed to global warming,
this author argued that the increase in APDI 
could be partially of anthropogenic origin.

However, \cite{Landsea_comment} noticed that, after applying the smoothing
procedure, Emanuel failed to drop the ending points, 
which were not averaged and were substantially larger than the 
smoothed series for the North Atlantic, creating the impression of a 
dramatic increase in PDI values.
Together with \cite{Gray_comment}, Landsea has also criticized
that Emanuel reduced the values of the wind speeds before the 1970's
by an excessive amount (up to more than 20 knots for the largest
values), in order to correct an overestimation in those values.
In fact,
recent research has shown that surface winds in major hurricanes
are stronger than previously assumed, and therefore, probably 
no correction at all is necessary \citep{Landsea_comment}.
The North Atlantic PDI series, with no correction, is plotted in Fig. \ref{fig2}.
Other problems with the accuracy of the records, 
particularly for the Eastern Hemisphere are pointed out
by \cite{Landsea_Science06}.

Many more articles have been devoted to these complex affairs, 
which unfortunately cannot be abstracted in this brief overview.
Before ending this section, let us just mention 
the work of 
\cite{Chan_comment},
%\cite{Hoyos},
\cite{Elsner_2006},
\cite{Emanuel_07},
\cite{Wu},
\cite{Swanson},
\cite{Elsner08}, 
\cite{Aberson}, and
\cite{Landsea_2010},
for its relation with the issues discussed here.

%graficas: num de TC's, hurris y major hurris

%Duraciones totales "no a la Gray" 

%PDI also

\section{PDI distribution and power-law fits}

All the studies summarized so far, counting numbers of storms,
total number of storm days, calculating NTC, ACE, or APDI, 
have paid attention to overall measures
of annual tropical-cyclone activity,
that is, they were trying to answer
the important question of comparing the characteristics
of different seasons, or longer periods.
But how different are the tropical cyclones from one season to another?
i.e., which are the features of individual tropical cyclones
in a given phase of activity, or under the influence
of a certain value of a climatic indicator?

The study of \cite{Webster_Science} contains a first
attempt in this direction, comparing the ratio between
tropical cyclones in a given Saffir-Simpson category
and tropical cyclones in all categories (1 to 5).
For global aggregated data it was found
that the proportion of category 4+5 events
increased from less than 20 \% in the early seventies
to about 35 \% after 2000,
with a corresponding decrease in the proportion
of category 1 storms, from more than 40 \% to 30 \%.
This is an indication that more major hurricanes 
(or typhoons) are present, in comparison,
but not that the storms are becoming more intense, 
individually
(i.e., there are more major ones in proportion, but their intensity
is not necessarily record-breaking).  
It is an open question in what part these results
could be an artifact due to the incompleteness
of the data for the earlier years \citep{Gray_comment,Klotzbach,Kossin}. 
In the North Atlantic, where the records are the best, 
these changes were much more modest, 
from a 20 \% of category 4+5 in 1975-1989 to 25 \% in 1990-2004.
In fact, what Webster and coworkers were really doing
was the calculation of the probability of hurricanes 
in a given category, and how this probability distribution changes 
from one period to another.

However, in order understand changes in 
the behavior of individual tropical cyclones
%to try to answer this, first of all, we need to know
we need to know, first of all,
which are the general properties of tropical cyclones, 
in other words, which is their ``unperturbed'' nature, 
and only after that we could study the influences 
of year-to-year variability on them.

This has been attempted by one of the authors and collaborators
\citep{Corral_hurricanes},
analyzing the statistics of the (individual-storm) PDI
for long periods of time.
As we have mentioned, the PDI is an approximation to dissipated
energy and is therefore perhaps the most fundamental characteristic
of a tropical cyclone.

In this way, the PDI for each tropical cyclone in the period and
basin considered was calculated, and the resulting probability density $D(PDI)$
was obtained, following its definition, as
$$
D(PDI) = \frac{\mbox{Prob}[ \mbox{value is in an interval around } PDI]}{\mbox{width of the interval}},
$$
where there is a certain freedom in the choice of the width of the interval (or bin)
\citep{Hergarten_book}
and Prob denotes probability, estimated by 
%%favorable cases divided by total number of cases.
relative frequency of occurrence.
Of course, normalization holds, $\int_0^\infty D(PDI) dPDI =1$.
For more concrete details on the estimation of $D(PDI)$,
see the supplementary information of \cite{Corral_hurricanes}.

\begin{figure}
 \noindent\includegraphics[width=20pc]{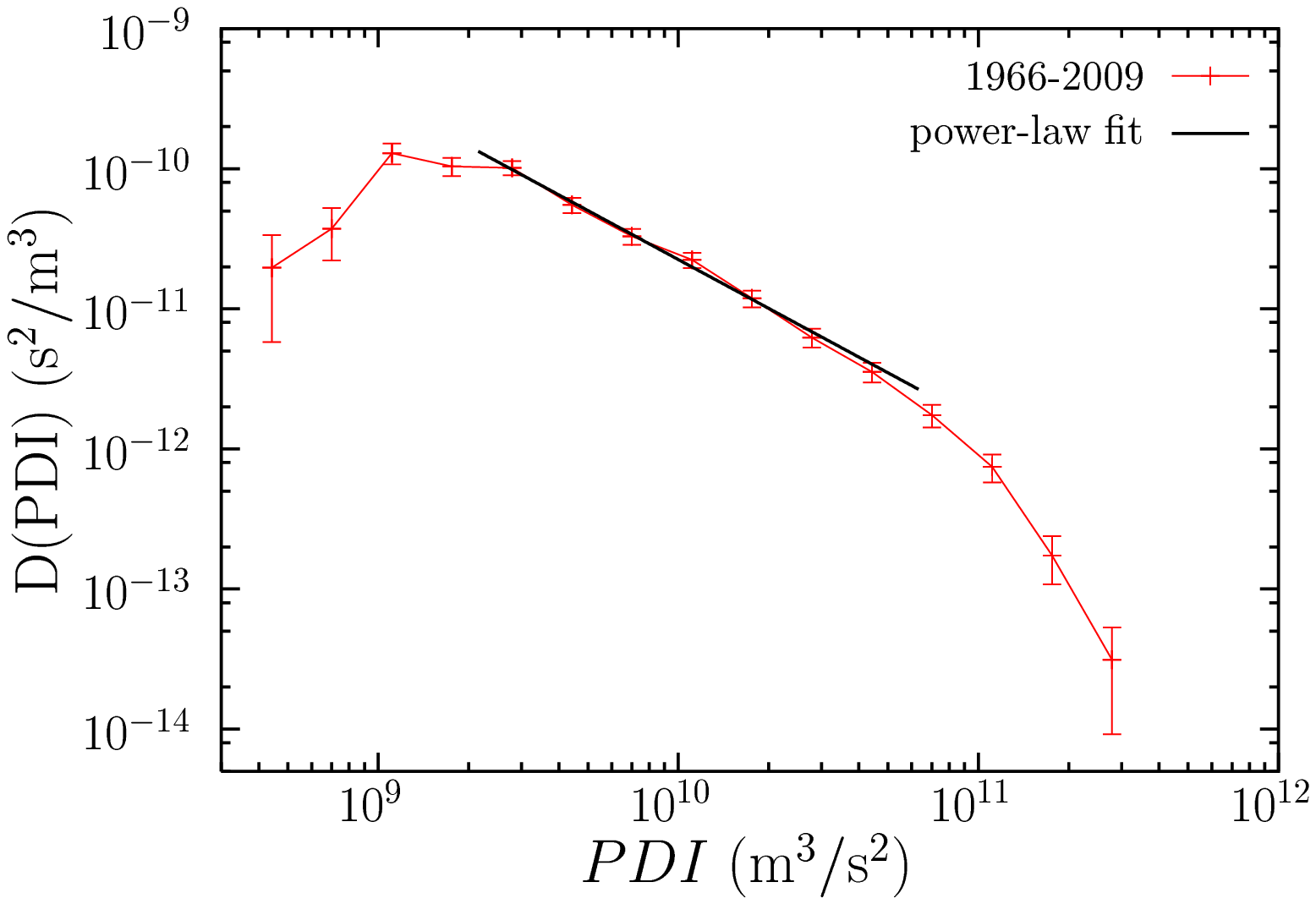}(a)\\
 \noindent\includegraphics[width=20pc]{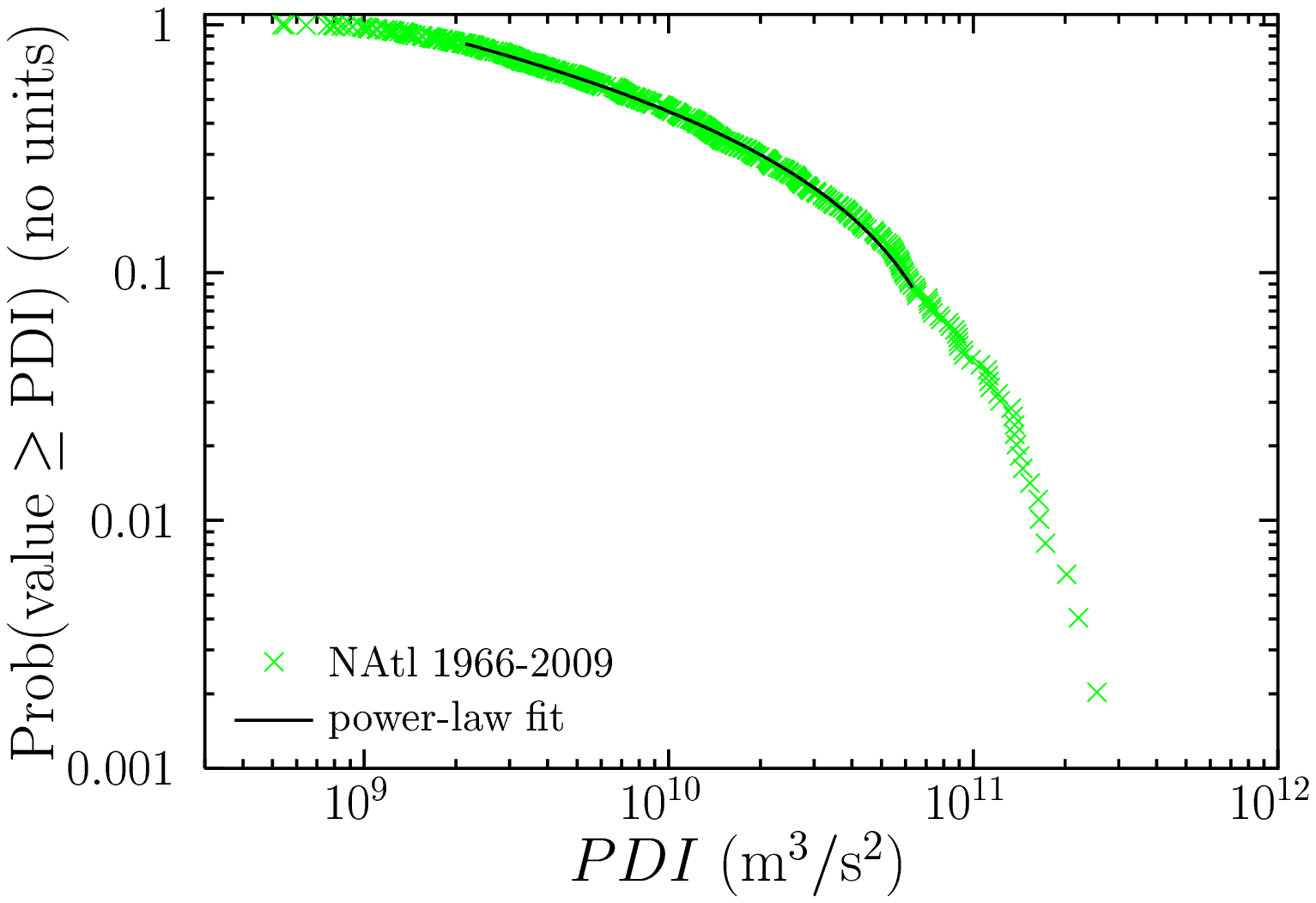}(b)
 \caption{
(a) Empirical probability density of 1966-2009 North-Atlantic 
individual tropical-cyclone PDI (494 events), together with a power-law fit
between $2.2 \cdot 10^9$ and $6.3 \cdot 10^{10}$ m$^3/$s$^2$
with an exponent $\tau=1.16$.
See \cite{Corral_hurricanes}.
(b) Same data and fit as in the previous panel but for the (complement of the) 
cumulative distribution function. 
Notice that the power-law distribution is no longer a straight line
in a log-log plot, due to the bending effect of the upper truncation,
which adds a constant to the power law
\citep{Burroughs_Tebbens}.
In this plot, the KS distance turns out to be proportional to the maximum difference
between the empirical cumulative density and its fit (rather small).
\label{fig3}}
 \end{figure}

The results for the 494 tropical cyclones of the North Atlantic
during the period 1966-2009
are shown in Fig. \ref{fig3}, in double logarithmic scale.
The data are the best tracks from NOAA's National Hurricane Center 
\citep{NOAA,NOAA_natl}; as we have seen, 1966 marks a year
from which the quality and homogeneity of the records is reasonable
(disregarding the possible overestimation of wind speeds
previous to 1970).
A straight line regime is apparent in Fig. \ref{fig3}(a), signaling the 
existence of a possible power-law behavior
over a certain range of PDI values,
$$
D(PDI) \propto \frac 1 {PDI^\tau},
\mbox{ for } a \le PDI < b,
$$
where $\propto$ indicates proportionality,
$\tau$ is the exponent and $a$ and $b$ are the truncation parameters.

Indeed, statistical tests support the power-law hypothesis.
The original reference \citep{Corral_hurricanes}
used a generalization of the fitting and testing procedure proposed by
\cite{Clauset}.
The key point is to find the optimum values of $a$ and $b$,
for which the following steps are followed: 

\begin{itemize}

\item Fix arbitrary initial values
of the truncation parameters $a$ and $b$,
with the only condition that $b/a > 10$,
for example.

\item Find the maximum likelihood estimation
of the exponent $\tau$,
by means of the maximization of
$$
\ell(\tau;a,b,G) = \ln \frac{\tau-1}{1-(a/b)^{\tau-1}}
-\tau \ln \frac G a -\ln a,
$$
or, alternatively, the solution of 
$$
\frac 1 {\tau-1} + \frac{a^{\tau-1}\ln(a/b)}{b^{\tau-1}-a^{\tau-1}}
-\ln\frac G a =0,
$$
with $G$ the geometric mean of the values comprised in the interval $[a,b)$.

\item Calculate %%(for the range $a \le PDI < b $) 
the Kolmogorov-Smirnov (KS) distance
between the empirical PDI distribution 
(defined only for the range $a \le PDI < b $)
and its power-law fit \citep{Press}.

\item Look for and select the values of $a$ and $b$
that minimize the Kolmogorov-Smirnov distance.
Select also the corresponding $\tau$ exponent.

\end{itemize}

Once the optimal values of $a$ and $b$, and from 
here the value of $\tau$, are found, 
a $p-$value quantifying the goodness of the fit
can be calculated just by simulating 
synthetic data sets, as close as possible 
to the empirical data (with the selected power-law exponent
$\tau$ between $a$ and $b$ and resampling the empirical
distribution outside this range).
The previous 4 steps are applied then to each of these
synthetic data sets (exactly in the same way as for the empirical data,
in order to avoid biases), which       
%The application of the generalized Clauset et al.'s
%procedure summarized above to each synthetic data set
yields a series of minimized Kolmogorov-Smirnov distances,
from which one can obtain the $p-$value, which is the 
survivor probability function of the synthetic KS distances
evaluated at the empirical value.
To be more precise,
$p$ is defined as the probability that, for true power-law distributed data, 
the KS distance is above the empirical KS value.
In this way, very small $p-$values are very unlikely under the 
null hypothesis, and one must conclude that the data are not
power-law distributed.
For technical details about the whole procedure, see the supplementary information
of \cite{Corral_hurricanes}.
The introduction of the (arbitrary) condition $b/a > 10$
is necessary because in the case of a double truncation
(from above and from below) several power-law regimes can
coexist in the data (in contrast to the original Clauset et al.'s case, $b\rightarrow \infty$),
and one needs a criterion to select the most appropriate one,
in our case that the range is high enough.

The results obtained from the application of this method to
the North-Atlantic 1966-2009 PDIs appear on table 1, 
%and a particular fit on Fig. \ref{fig3},
and they show that there is no reason to reject the power-law hypothesis
for a certain set of values of $a$ and $b$.
Among the several outcomes of the method, depending on the conditions
imposed, a good choice is the one with the values of $a$ and $b$
around $2 \cdot 10^9$ and $6 \cdot 10^{10}$ m$^3/$s$^2$,
with a resulting exponent $\tau =1.16$ and a $p-$value about 40 \%,
see Fig. \ref{fig3}.
In general, one needs to balance the tendency to select low $b/a$ ratios,
which yield high $p-$ values, with the tendency to larger ratios and lower $p$.

In \cite{Corral_hurricanes} it was checked that the behavior of
the PDI distribution is very robust, showing very little variation 
for different time periods, or if 
extratropical, subtropical, wave and low stages are not taken into account, 
or if U.S. landfalling tropical cyclones are removed from the record,
or speeds previous to 1970 are reduced by an amount of 4 m/s
\citep{Landsea93}.
In addition, the power-law behavior was also found in other basins:
the Northeastern Pacific, the Northwest Pacific and the Southern
Hemisphere as a whole (the North Indian ocean was not analyzed due
to the low number of events there and the lack of a complete satellite monitoring 
until recently \citep{Kossin}). The resulting exponents were $\tau=1.175$, $0.96$, and $1.11$,
respectively, $\pm 0.05$, in the worst case.

Further, it has been pointed out that the power-law PDI
distribution could be a reflection of the criticality
of hurricane occurrence \citep{Corral_Elsner},
in the same way as for earthquakes, forest fires, rainfall, sandpiles, etc.
\citep{Bak_book,Turcotte_book,Jensen,Sornette_critical_book,Christensen_Moloney}.
In fact, tropical-cyclone criticality 
could be only a part of the criticality
of atmospheric convection \citep{Peters_np,Peters_ijmpb}.
The parallelisms between tropical-cyclone occurrence
and self-organized criticality are discussed in depth in 
\cite{Corral_Elsner}.
The implications of this finding for the predictability
of the phenomenon would be rather negative, 
in the sense that this kind of processes
are characterized by an inherent unpredictability,
although the differences with deterministic chaos 
are worth of being investigated in the future.

Nevertheless, the previous statistical method was not found to be fully satisfactory
when applied to a different data set \citep{Corral_nuclear}.
The problem lies in the selection criterion for
$a$ and $b$.
First, it can provide non optimal values of them, 
which would lead to the rejection of the null hypothesis
in cases in which is true.
Second, it has a strong tendency to underestimate
the upper limit $b$ (if $\tau > 1$).

Another method, with a distinct selection procedure for $a$ and $b$,
was introduced by \cite{Peters_Deluca}.
The steps are:
\begin{itemize}

\item Fix arbitrary initial values
of the truncation parameters $a$ and $b$,
with no restrictions.

\item Find the maximum likelihood estimation
of the exponent $\tau$
(by means of the same formula
as for the previous method).

\item Calculate the Kolmogorov-Smirnov distance
between the empirical PDI distribution (defined only for the range $a \le PDI < b $)
and its power-law fit.

\item
Calculate the $p-$value of the fit,
which we denote by $q$, by simulations
of synthetic power-law data sets, 
with exponent $\tau$
(the simulation of the distribution
outside the interval $[a,b)$ 
is not necessary in this case).

\item Look for and select the values of 
$a$ and $b$ which maximize the logarithmic range,
$b/a$, under the condition, let us say, $q> 20 \%$.

\end{itemize}

In the original reference by Peters et al.
the number of data in between $a$ and $b$, $N_{ab}$, 
was the quantity which was maximized, 
rather than the log-range $b/a$.
As in our case the power-law exponent is very close
to one (for which there is the same number of data for each order of magnitude), 
there should be no relevant difference
between both procedures;
nevertheless, the we think the present choice is more
appropriate for tails with larger exponents.
Of course, setting the threshold $q$ equal to 20 \%
is arbitrary, and we will have to check the effect
of changing this value.

One has to take into account that the
$q$ value calculated in this method is not the true $p-$value
of our fitting procedure.
The former is calculated for fixed, or known, $a$ and $b$
values, whereas the latter should take into account that
both parameters are optimized.
Comparing with the previous method it seems that
the change between both $p-$values can be 
around a factor 2 or 3 (so caution is necessary at this point).
The results of this method for the same North Atlantic data used in table 1
are displayed now in table 2.
Indeed, a comparison between both tables show that the results are
consistent, if one takes into account the difference between $p$ and $q$.
In summary, the existence of  a power law over a range larger than one decade
is well supported by the statistical tests.

\begin{table}
\begin{tabular}{r|r|r|r|r|r|c|r|r|r|}
 $R_{min}$ & $a \, ($m$^3/$s$^2)$ & $b \, ($m$^3/$s$^2)$&  $b/a$ & $N_{ab}$  & 
$C N_{ab}/N$ & $ \tau  \pm \sigma_\tau$ & $d$ & $d\sqrt{N_{ab}}$ & $p-$value 
%% \\ \hline&&&&&
\\ \hline
%%%%%%%%%%%%%%%%%%%%%%%%%%%%%%%%%%%%%%%%%%%%%%%%%%%%%%%%%%%%%%%%%%%%%%%%%%
%                         nuevo, variando b/a ,  100 simus
%%%%%%%%%%%%%%%%%%%%%%%%%%%%%%%%%%%%%%%%%%%%%%%%%%%%%%%%%%%%%%%%%%%%%%%%%%
 % rangemin fit       1.100000023841858
 % amin  2.690000000000000E+008
 % bmin  5.088000000000000E+011
 % points dec      30.000000000000000
 % nsim        100
%%%        1.1 &   1.8 $10^9$ &   1.7 $10^{10}$ &        9.3 &         270 &   0.36     &        1.02 $\pm$  0.09 &        2.4 $10^{-2}$ &   0.39 &       57 $\pm $       5 \% \\
        2 &   1.8 $10^9$ & 
  1.7 $10^{10}$ &        9.3 &         270 & 
  0.36     &        1.02 $\pm$
  0.09 &        2.4 $10^{-2}$ & 
  0.39 &       57$\pm $5 \% \\
%%%        5. &   1.8 $10^9$ &   1.7 $10^{10}$ &        9.3 &         270 &   0.36     &        1.02 $\pm$  0.09 &        2.4 $10^{-2}$ &   0.39 &       58 $\pm $       5 \% \\
       10 &   2.2 $10^9$ & 
  6.3 $10^{10}$ &       29.3 &         371 & 
       8.6 &        1.16 $\pm$
  0.05 &        2.5 $10^{-2}$ & 
  0.47 &       39$\pm $5 \% \\
%%%       20. &   2.2 $10^9$ &   6.3 $10^{10}$ &       29.3 &         371 &        8.6 &        1.16 $\pm$  0.05 &        2.5 $10^{-2}$ &   0.47 &       47$\pm $       5 \% \\
%%%       30. &   2.2 $10^9$ &   6.8 $10^{10}$ &       31.6 &         375 &       10.6 &        1.17$\pm$  0.05 &        2.5 $10^{-2}$ &   0.48 &       54$\pm $       5 \% \\
%%%      40. &   1.8 $10^9$ &   7.9 $10^{10}$ &       43.0 &         397 &        8.3 &        1.16 $\pm$  0.05 &        2.9 $10^{-2}$ &   0.58 &       15$\pm $       4 \% \\
       50 &   1.8 $10^9$ & 
  9.3 $10^{10}$ &       50.1 &         405 & 
      12.1 &        1.175$\pm$%
  0.04 &        3.1 $10^{-2}$ & 
  0.63 &       14$\pm $3  \% \\
       70 &   1.7 $10^9$ & 
  12.6 $10^{10}$ &       73.6 &         421 & 
      23.3 &        1.205$\pm$%
  0.04 &        4.2 $10^{-2}$ & 
  0.87 &        2$\pm $1 \% \\
      100 &   1.4 $10^9$ & 
  14.7 $10^{10}$ &      108.0 &         452 & 
      15.4 &        1.19 $\pm$
  0.04 &        5.0 $10^{-2}$ & 
       1.05&   0  \% \\
% rangemin fit      20.000000000000000
 % amin  2.690000000000000E+008
 % bmin  5.088000000000000E+011
 % points dec      30.000000000000000
 % nsim        100
\hline
%%\footnote{For the period 1966-2007}
     20 &   2.7 $10^9$ & 
  6.3  $10^{10}$ &       23.3 &         328 & 
      18.3 &        1.19 $\pm$
  0.06 &        2.4 $10^{-2}$ & 
  0.43 &       67$\pm $5 \% \\
\end{tabular}
\caption{
Parameters of the maximum likelihood estimation and the KS test
for the PDI of the 494 North Atlantic tropical cyclones
of the period 1966-2009,
using the generalization of the method of \cite{Clauset}
introduced by \cite{Corral_hurricanes}.
The condition $b/a > R_{min}$ is imposed.
$N_{ab}$ refers to the number of tropical cyclones with PDI value
between $a$ and $b$;
$C N_{ab}/N$ is the constant of the power law that fits the empirical
distribution between $a$ and $b$ when the latter is normalized
from 0 to $\infty$, its units are (m$^3/$s$^2$)$^{\tau-1}$.
The uncertainty of the exponent can be obtained from its standard
deviation, $\sigma_\tau$ \citep{Aban},
which, in the limits of large $N_{ab}$ and $b \gg a$, is given by
$\sigma_\tau \simeq (\tau-1)/\sqrt {N_{ab}}$.
$d$ is the minimized KS distance,
and $d\sqrt{N_{ab}}$ a rescaling of that distance
that should be independent on ${N_{ab}}$,
for comparison purposes.
$a$ and $b$ are determined with a resolution of 30 points per order of magnitude.
The $p-$values are calculated from 100 Monte Carlo simulations.
The last row corresponds to the period 1966-2007,
analyzed by \cite{Corral_hurricanes},
where an erratum was present in the value of $C N_{ab}/N$.
\label{table1}
}
\end{table}

\begin{table}
\begin{tabular}{r|r|r|r|r|r|c|r|r|}
 $q_{min}$ & $a \, ($m$^3/$s$^2)$ & $b \, ($m$^3/$s$^2)$&  $b/a$ & $N_{ab}$  & 
$C N_{ab}/N$ & $ \tau  \pm \sigma_\tau$ & $d$ & $d\sqrt{N_{ab}}$ %%%%% & $q-$value 
%% \\ \hline&&&&&
\\ \hline
   99 \%
  &   14.7 $10^{9}$ &   6.8 $10^{10}$ &    5 &     134
  &       4.28
  &        1.13 $\pm$  0.20 
  &        3.5 $10^{-2}$ &        0.41
% &      100 $\pm $  0.E+000 \% 
\\
   90 \%
  &   2.2 $10^9$ &   6.3 $10^{10}$ &    29 &     371
  &       8.59 
  &        1.16 $\pm$  0.05 
  &        2.5 $10^{-2}$ &        0.47 
% &       94 $\pm $       2.374868417407584 \% 
\\
   50 \%
  &   2.0 $10^9$ &   9.3 $10^{10}$ &   46 &      395
  &       12.1 
  &        1.175$\pm$  0.05
  &        3.0 $10^{-2}$ &        0.59 
% &       54 $\pm $       4.983974317750845 \% 
\\
%%%%   75 \%  &   2.2 $10^9$ &   7.3 $10^{10}$ &    34 &     381  &       19.4   &        1.17 $\pm$  0.05   &        2.6 $10^{-2}$ &        6.3   &       88 $\pm $       3.249615361854384 \% \\
   20 \%
  &   1.7 $10^9$ &   10.0 $10^{10}$ &  58 &       414
  &       11.9
  &        1.175$\pm$  0.04
  &        3.7 $10^{-2}$ &        0.76
% &       28 $\pm $       4.489988864128730 \% 
\\
   10 \%
  &   1.7 $10^9$ &   14.7 $10^{10}$ &  86 &      428
  &       30.1 
  &        1.22 $\pm$  0.04
  &        4.3 $10^{-2}$ &        0.90 
% &       12 $\pm $       3.249615361854384 \% 
\\
\end{tabular}
\caption{
Same as the previous table but using the fitting procedure
of \cite{Peters_Deluca}, which maximizes the log-range 
under the condition $q >q_{min}$,
where $q$ denotes the $p-$value for fixed $a$ and $b$.
\label{table2}
}
\end{table}

\section{Gamma distribution of PDI}

So far, we have established that a ``significant''
range of the PDI probability density
can be described as a power law.
Deviation at small values ($PDI < a$)
is justified through the fact that the
best-track records are incomplete --
this is obvious, as tropical depressions
are deliberately excluded from the database.
Inclusion of tropical depressions in the analysis of the Northwestern Pacific
(which are easily available, in contrast to tropical depressions at 
the NHC best-track records) 
enlarges somewhat the power-law range,
although the coverage of 
tropical depressions in that basin is far from exhaustive
\citep{Corral_hurricanes}.

The rapid decrease of the PDI density above the value of $b$
has been explained as a finite size effect:
tropical-cyclones cannot become larger (in terms of PDI)
because they are limited by the finiteness of the basin;
either they reach 
extratropical regions (cold conditions)
or they reach land, in which cases they are dissipated
as they are deprived from their warm-water energy source.
In this way, it has been shown how the tracks of most events with $PDI > b$
are affected by the boundaries of the basin,
and a finite-size scaling analysis showed how a reduced 
area (limiting it in terms of longitude) moved the cutoff to smaller values, 
with the cutoff defined roughly as the point in which the PDI 
probability density clearly departs from the power law
\citep{Corral_hurricanes}.

Interestingly, it has also been found 
in the previously mentioned reference
that the season-averaged tropical SST 
of the North Atlantic has an influence on the PDI values similar
to a finite size effect.
Years with high SST lead to a larger cutoff, just the opposite as
years with low SST, but keeping nearly the same value of the power-law exponent
(the same has been found also for the Northeastern Pacific).
So, the power-law is a robust feature of the PDI distribution, but it is not telling us 
anything about the influence of other factors (finiteness of the basins and SST);
rather, it is the cutoff, which is not properly defined yet, which carries this information.

In order to overcome this deficit, a function modeling the PDI distribution
that includes the tail seems appropriate.
Experience with finite size effects in critical phenomena
(taking place in continuous phase transitions or in avalanche-evolving systems)
suggests a simple exponential factor for the tail,
%suggests that a simple exponential tail is a reasonable choice 
%\citep{Christensen_Moloney},
%seguro?
and therefore a gamma-like distribution for the whole domain
(excepting the incompleteness for small values),
$$
D(x) = \frac 1 {c \Gamma(-\beta,a/c) }
\left(\frac{c}{x}\right)^{1+\beta} e^{-x/c},
\mbox{ for } x \ge a, 
$$
where $x$ plays the role of the PDI
(only for aesthetic reasons),
$c$ represents now the cutoff, 
which is a scale parameter,
and $\beta$ is a shape parameter,
with the power-law exponent equal to $1+\beta$.
If $a=0$, $\beta$ has to be negative, 
in order to avoid the divergence of the integral of $D(x)$;
however, for $a > 0$, there is no restriction on $\beta$.
The (complement of the unrescaled) incomplete gamma function is
$$
\Gamma(\gamma,z_o) =\int_{z_o}^\infty z^{\gamma-1} e^{-z} dz
$$  
\citep{Abramowitz},
with $\gamma>0$ for $z_o=0$ but unrestricted
for $z_o > 0$.
Unfortunately, some of the numerical routines we will use
are not defined for $\gamma\le 0$,
so, we will need to transform
$$
\frac 1 {c \Gamma(-\beta,a/c) }=
\frac \beta c \left[ \left( \frac c a \right)^\beta e^{-a/c} - \Gamma(1-\beta,a/c)  \right]^{-1}
$$
(integrating by parts),
where, for $\beta< 1$ the incomplete gamma function on the right hand side
can be computed without any problem.
An alternative approach for the evaluation of the cutoffs
would have been to compute the moment ratios used by \cite{Peters_Deluca}.

The reason to model $D(PDI)$ by means of a gamma distribution
is due to the fact that one can heuristically understand 
the finiteness of a system as introducing a kind of effective correlation 
length. 
Then, it is well known that, in the simplest cases, correlations  
enter into probability distributions under an exponential form
\citep{Zapperi_branching}.

In contrast to the method of fit and goodness-of-fit 
test explained above and to other previous work
\citep{Corral_test}, in order to avoid difficulties, 
we perform a simple fitting procedure here,
acting directly over the estimated probability density.
The value of the minimum limit $a$ is selected 
from information coming from the power-law fits (tables 1 and 2)
complemented by visual inspection
of the plots of the empirical probability density.
A reasonable assumption is $a=2 \cdot 10^9$ m$^3/$s$^2$
(note that the empirical $D(PDI)$ is discretized).

The fit of the shape and scale parameters, $\beta$ and $c$,
is performed using the
nonlinear least-squares Marquardt-Levenberg algorithm
implemented in {\tt gnuplot}, applied to the logarithm
of the dependent variable (i.e., it is the logarithm of the model
density which is fit to the logarithm of empirical density);
further, the parameter introduced in the algorithm is not
$c$ but $c'=\ln c$
(i.e., we write $c = \exp(c')$).
Moreover, it is necessary to correct that the empirical density is
normalized from 0 to $\infty$ (i.e., not truncated,
all events contribute to the normalization),
whereas the model density goes from $a$ to $\infty$.
In order that one can fit the other, a factor $N_a/N$ 
must multiply the fitting density
(or, equivalently, divide the empirical one 
by the same factor), where $N_a$ is the number of data points (tropical cyclones)
fulfilling $x\ge a$.
Due to the discretization of the empirical density,
in practice, $a$ has to coincide with the lower limit of a bin
of the density, or, alternatively, 
$N_a$ has to be redefined accordingly.
%We have a certain error source here, as the 
%discretization of the density implies that the number of
%points it contains for $x\ge a$ is not exactly $N_a$.
%This is only true if $a$ coincides with the lower limit of a bin.
%In any case, if the bin width is not very large, 
%the errors will be small.
%The good fits we obtain certify that we do not need to care
%about this point.

\begin{figure}
 \noindent\includegraphics[width=20pc]{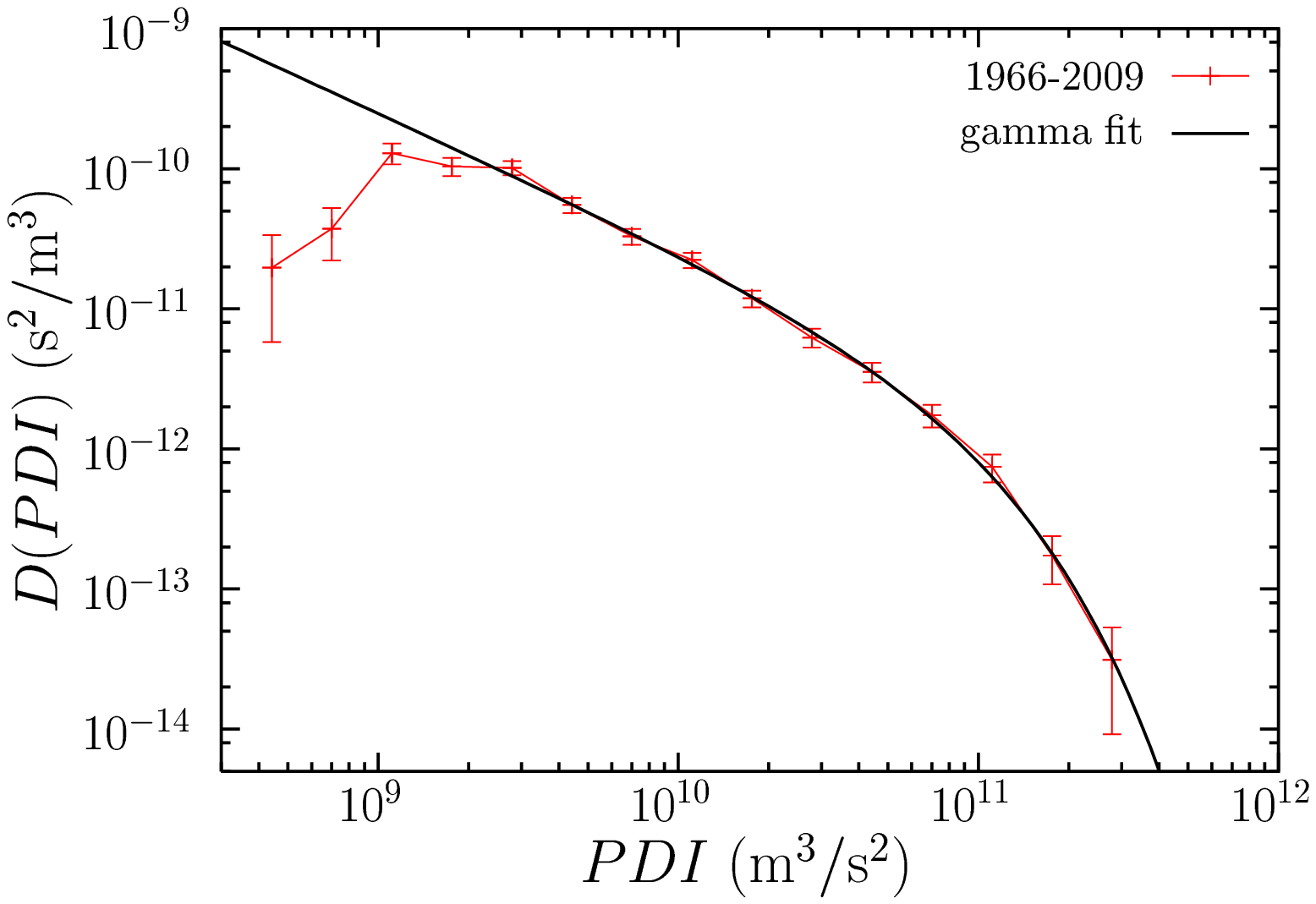}
 \caption{
Same North-Atlantic tropical-cyclone data than in the previous figure, 
but with a gamma fit for $PDI\ge 2 \cdot 10^{10}$ m$^3/$s$^2$. 
The exponent is $\tau=1+\beta=0.98$ and the scale parameter
$c=8.1 \cdot 10^{10}$ m$^3/$s$^2$.
The fit seems reasonably good for 2 orders of magnitude.
\label{fig4}}
 \end{figure}

The resulting fit
for the 1966-2009 PDIs of the North Atlantic 
is displayed on Fig. \ref{fig4},
and the 
obtained parameters
are %%%$a=2 \cdot 10^9$ m$^3/$s$^2$
$\tau=1+\beta=0.98 \pm 0.03$ and
$c=(8.1 \pm 0.4) \cdot 10^{10}$ m$^3/$s$^2$,
when the empirical density is estimated with 5 bins per order of magnitude
(and the position of those bins is the one in the figure). 
In contrast with the method used for power-law fits,
the results here depend on the width of the intervals
of the empirical probability density (the number of bins or
boxes per order of magnitude), and their position in the $x-$axis.
We have found that, in general, the influence of these factors
on the parameters is more or less within the error bars given above.

Note that the exponent $\tau$ was larger than 1
for the (double truncated) power law,
whereas now $\tau=1+\beta$ turns out to be slightly below 1.
The difference is small, but, more importantly, 
both exponents represent different things,
because we are fitting different functions.
If the true behavior of $D(PDI)$ were a gamma distribution
(which is in practice impossible to know),
it is expected that a pure power law
(over a certain range) would yield a larger
(steepest) exponent, due to the bending effect of
the exponential factor.

On the other hand, a gamma distribution with $1+\beta \le 1$
cannot be extrapolated to the case in which the 
cutoff is infinite and cannot describe vanishing finite size effects,
as it would not be normalizable.
If we enforce $1+\beta \ge 1$ in the fit, 
we get 
$1+\beta=1.00 \pm 0.03$
and
$c=(8.3 \pm 0.4) \cdot 10^{10}$ m$^3/$s$^2$,
which is not only a visually satisfactory fit
but undistinguishable from the previous one.
So, there is a tendency of the exponent to be
smaller than one, but values slightly above one
are equally acceptable.

\section{Sensitivity of the tail of the PDI distribution to SST
%%%and other climatic variables
}

We now investigate 
how the splitting of the PDI distribution 
into two parts, one for warm SST years
and another for colder years,
influences the shape and scale of
the resulting distributions. 
%(through the values of the parameters $\beta$ and $c$).
First, monthly SST with one degree spatial resolution are
averaged for the tropical North Atlantic
($90^\circ$ to $20^\circ$W, $5^\circ$ to $25^\circ$N)
during the hurricane season (June to October).
This yields a single SST number for each year or season
\citep{Webster_Science}, and in this way, 
values above the 1966-2009 mean
are considered to denote warm years, 
and the opposite for cold years
(notice that the comparison is not done relative
to a different, longer period).
The average SST for warm years turns out to be 0.49 $^\circ$C larger
than that of cold years.
The SST data are those of the Hadley Centre, UK Meteorological Office 
\citep{SST,SST_webnew}.

\begin{figure}
 \noindent\includegraphics[width=20pc]{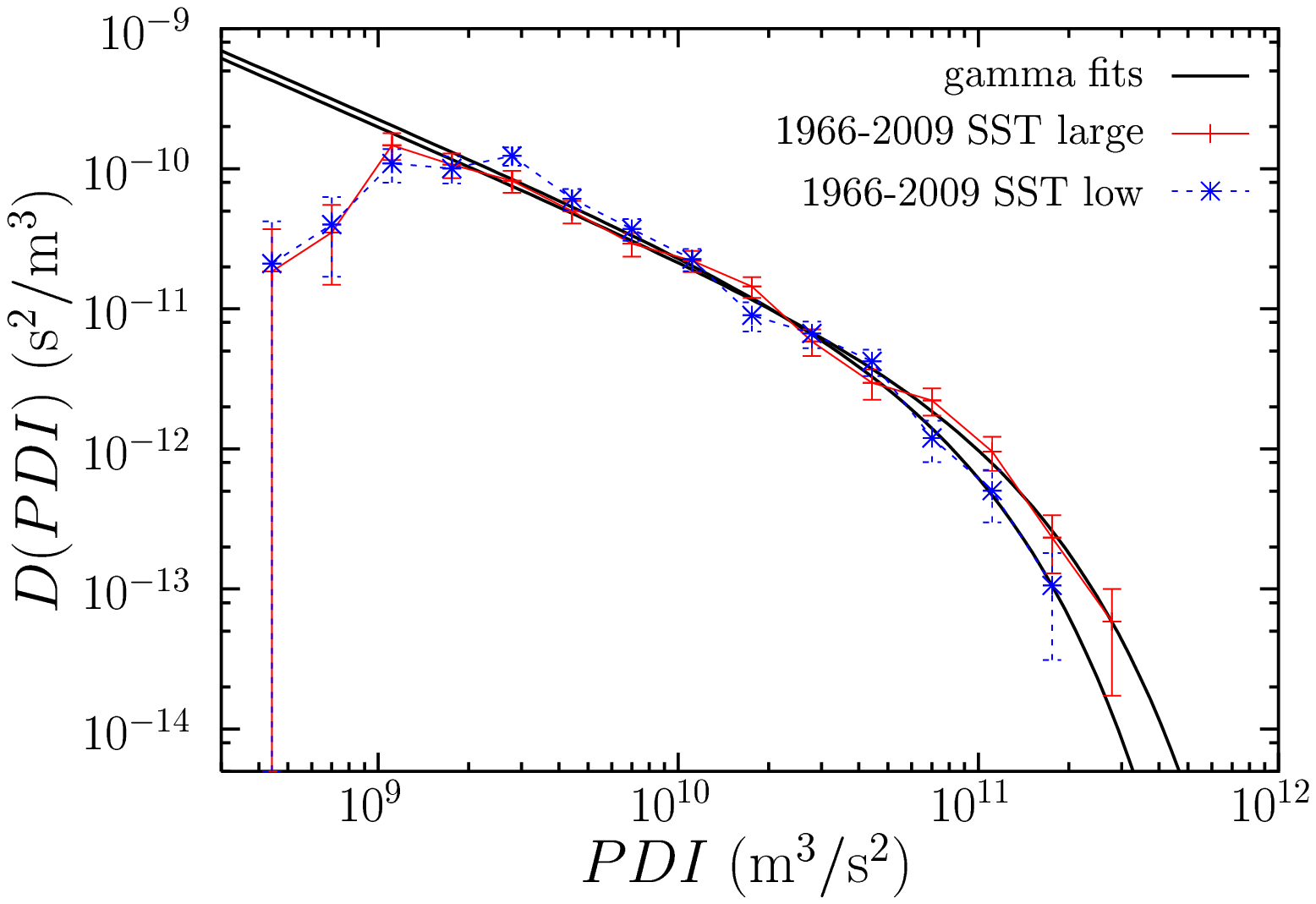}
 \caption{
Same North-Atlantic PDI data than for the two previous figures, 
but separated into two distributions, one for years with SST above the mean
(large in the label) and another for below (low). 
The gamma fits, enforcing a constant exponent $1+\beta=0.93$ (see text),
show an increase in the scale parameter $c$ around 50 \% between
cold and warm SST years.
\label{fig5}}
 \end{figure}

The resulting PDI distributions are displayed on Fig. \ref{fig5}, which shows
that the largest values of the PDI correspond to high SST years.
If we compare the mean values of the PDI, the increase is around 40 \%, 
i.e., 
$\langle PDI \rangle_{warm}/\langle PDI\rangle_{cold}\simeq 1.42 \pm 0.19$, 
which is significantly larger than 1 (at the 99 \% confidence level, see table 3).
%The reason is that, although the increase is huge, 
%the uncertainty is also very large, as
%the PDI is broadly distributed.

The gamma fit also does a good job here, 
with an outcome, for high SST, 
$1+\beta= 0.94 \pm 0.06$ and $c=(9.6 \pm 1) \cdot 10^{10}$ m$^3/$s$^2$,
taking $a=2 \cdot 10^{9}$ m$^3/$s$^2$.
For low SST, visual inspection of the plot of the density suggests
that it is more appropriate to raise the value of $a$ to $3\cdot 10^{9}$ m$^3/$s$^2$
and then,
$1+\beta= 0.98 \pm 0.11$ and $c=(6.5 \pm 1) \cdot 10^{10}$ m$^3/$s$^2$.
This yields a $50 \pm 27 $\% increase in the value of the cutoff, 
which is significantly larger than zero if one assumes
normality for the $c$ parameters.
Note that we do not expect differences between the ratios of the cutoffs
and the ratios of the means, due to the fact that the power-law
exponent $1+\beta$ turns out to be smaller than 1,
and then the mean scales linearly with the cutoff,
see the supplementary information of \cite{Corral_hurricanes}.

If, in order to make the comparison of $c$ in the same conditions
between warm and cold years,
we fix $a=3 \cdot 10^9$ m$^3/$s$^2$ and $1+\beta=0.93$ (this is the value obtained
in this case for the whole dataset, with $c=(7.8 \pm 0.4) \cdot 10^{10} $ m$^3/$s$^2$),
then we get $c=(9.6 \pm 0.5)\cdot 10^{10} $  m$^3/$s$^2$ and
$(6.1 \pm 0.4)\cdot 10^{10} $  m$^3/$s$^2$,
for high and low SST years, respectively,
which is not essentially different to the above case,
but has smaller errors. 
This leads to
$c_{warm}/c_{cold}=1.57 \pm 0.13$,
i.e., more than a 50 \% increase in the largest values of the PDI
for an increase in the SST equal to 0.49 $^\circ$C,
%(which is less than a 0.2 \% in absolute temperature)
which seems to indicate a high sensitivity of the
extreme (cutoff) PDIs to the warming of the sea surface.
The fits are displayed in Fig. 5.

A further step is to separate the years not into two groups,
warm and cold, but into 3, very warm, intermediate and very cold
(in relative terms), 
and compare only the years with ``extreme'' temperatures.
Taking the thresholds as the mean SST plus half its standard
deviation and the mean minus half the standard deviation
we get a ratio of mean PDIs equal to 
$1.65 \pm 0.25$, for an increase in mean SST about $0.65^\circ$C,
see table 3.
Thus, the splitting of the distributions is more pronounced
for ``extreme'' SST.

\begin{table}
\begin{tabular}{l|l|l|l|l|l|l|}
                            & SST           & MEI           & NAO           & AMO           & $n_{TC}$      & $n_{H}$ \\ \hline
variable difference         & 0.49$^\circ$C & 1.41          & 0.79           & 0.39         & 6.55 & 4.45\\
$\langle PDI\rangle_{high}$ & 2.55$\pm$0.24 & 2.09$\pm$0.23 & 2.11$\pm$0.20 & 2.50$\pm$0.24 & 2.39$\pm$0.22 & 4.27$\pm$0.36 \\ %% & 2.71$\pm$0.25 \\
$\langle PDI\rangle_{low}$  & 1.79$\pm$0.17 & 2.29$\pm$0.21 & 2.26$\pm$0.22 & 1.85$\pm$0.18 & 1.92$\pm$0.21 & 2.83$\pm$0.26 \\ %% & 1.58$\pm$0.16 \\
$\langle PDI\rangle$ ratio  & 1.42$\pm$0.19 & 0.91$\pm$0.13 & 0.93$\pm$0.13 & 1.35$\pm$0.18 & 1.25$\pm$0.18 & 1.51$\pm$0.19 \\ %% & 1.71$\pm$0.32
$\langle PDI\rangle$ 
difference                  & 0.76$\pm$0.30 &-0.20$\pm$0.31$^*$ &-0.15$\pm$0.30$^*$ & 0.65$\pm$0.30 & 0.48$\pm$0.30$^*$ & 1.45$\pm$0.45 \\
\hline
variable difference                & 0.65$^\circ$C & 1.92          & 1.15           & 0.52         & 9.47 & 6.32 \\
$\langle PDI\rangle_{very\, high}$ & 2.70$\pm$0.29 & 1.39$\pm$0.17 & 2.43$\pm$0.27 & 2.66$\pm$0.31 & 2.53$\pm$0.29 & 4.26$\pm$0.41 \\ %%& 2.78$\pm$0.29 \\
$\langle PDI\rangle_{very\, low}$  & 1.64$\pm$0.18 & 2.10$\pm$0.24 & 2.22$\pm$0.30 & 1.56$\pm$0.17 & 1.60$\pm$0.22 & 2.51$\pm$0.31 \\ %%& 1.31$\pm$0.17 \\
$\langle PDI\rangle$ ratio         & 1.65$\pm$0.26 & 0.66$\pm$0.11 & 1.09$\pm$0.19 & 1.70$\pm$0.27 & 1.58$\pm$0.28 & 1.70$\pm$0.26 \\ %%& 2.12$\pm$0.50 \\
$\langle PDI\rangle$
difference                         & 1.06$\pm$0.34 &  -0.71$\pm$0.30 & 0.21$\pm$0.40$^*$ & 1.09$\pm$0.35 & 0.93$\pm$0.36 & 1.75$\pm$0.51  \\
%\hline
%%& & & & & & {\tiny $^*$Only for hurricanes.}                               \\
\end{tabular}
\caption{
Top half:
Mean value of the PDI for high and low values
of seasonal averaged values of the SST, MEI, NAO, and AMO, 
together with the annual number $n_{TC}$ of tropical cyclones,
and the annual number of hurricanes, $n_H$,
for the North Atlantic in the period 1966-2009.
For the latter case the averages are performed
only for hurricanes (otherwise we could get artificial 
larger ratios due to the fact that years with high number of hurricanes
and less tropical storms would yield higher values of the PDI).
The difference of the variables between the mean for high and low years is shown, 
as well as the ratio and the difference between the mean values of the PDI.
%and the difference divided by its uncertainty
%(in order to compute statistical significance).
An asterisk denotes non-significant differences (at the 95 \% level).
Units of $\langle PDI\rangle$ and its differences are $10^{10}$ m$^3/$s$^2$. 
The uncertainty of the mean PDI is the standard deviation of the PDI
divided by the square root of the number of data.
The relative uncertainty of the ratio is the square root of the sum of square of the
relative errors of each value of the means.
Bottom half:
The same but for values of the SST and other variables
above the mean plus one half of the standard deviation
(very high) and 
below the mean minus one half of the standard deviation
(very low).
\label{table3}
}
\end{table}

\section{Influence of El Ni\~no, the NAO, the AMO, and the number of storms
on the PDI}

We can proceed in the same way for other climatic variables,
or indices.
Considering the El Ni\~no southern oscillation (ENSO), the
multivariate ENSO index (MEI) separates warm phases 
of the tropical Eastern Pacific (MEI $>0$)
from cold phases (MEI $<0$), with high values of the MEI
associated to El Ni\~no and low values to La Ni\~na 
\citep{MEI2}.
It has been proved that El Ni\~no phenomenon is anti-correlated with 
hurricane activity in the North Atlantic,
in such a way that the presence of El Ni\~no
partially suppresses this activity \cite{Gray_84}, 
which has been detected even in 
the record of hurricane economic losses in the U.S.
\citep{Pielke_ElNino}.
Using data from \cite{MEI_webnew},
and averaging for each season the months from May/June to September/October,
we can distinguish between positive and negative phase seasons, 
and from here obtain the corresponding PDI distributions.

We do the same for the NAO and AMO 
indices (North Atlantic oscillation and Atlantic multidecadal oscillation, respectively,
averaging the indices from June to October),
as well as for the annual number of tropical cyclones, $n_{TC}$,
and the annual number of hurricanes, $n_H$.
In the latter case, only hurricanes are accounted in 
the average of the PDI.
High values of MEI, NAO, and AMO refer to just positive values of the indices,
whereas high values of $n_{TC}$ and $n_{H}$ denote values above the
1966-2009 mean (in the same way as for the SST),
and conversely for low values of the variables.
Very high values correspond in all cases to values above
the mean plus one half of the standard deviation
(and conversely for very low values).

\begin{figure}
 \noindent\includegraphics[width=20pc]{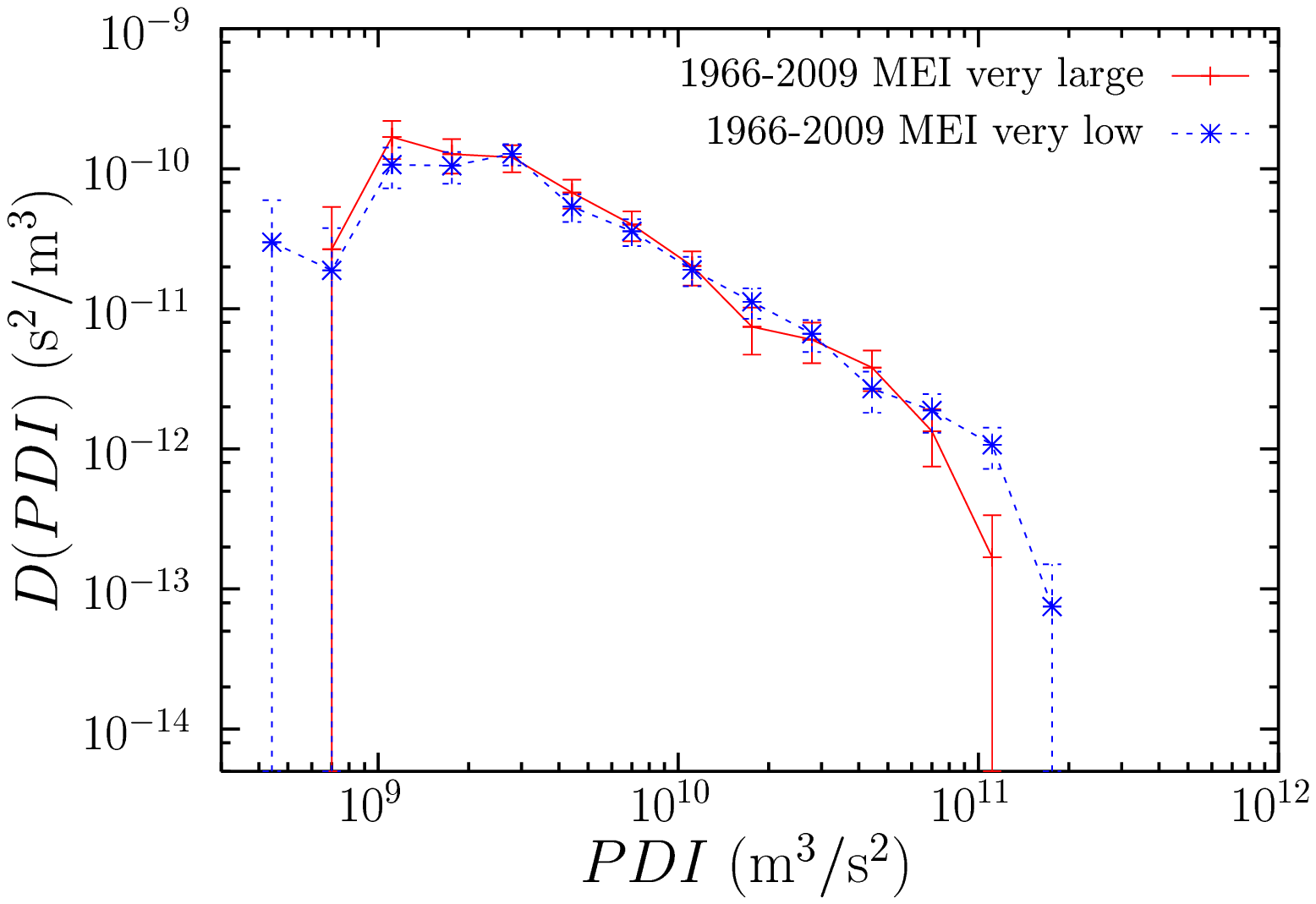}
(a) \\
 \noindent\includegraphics[width=20pc]{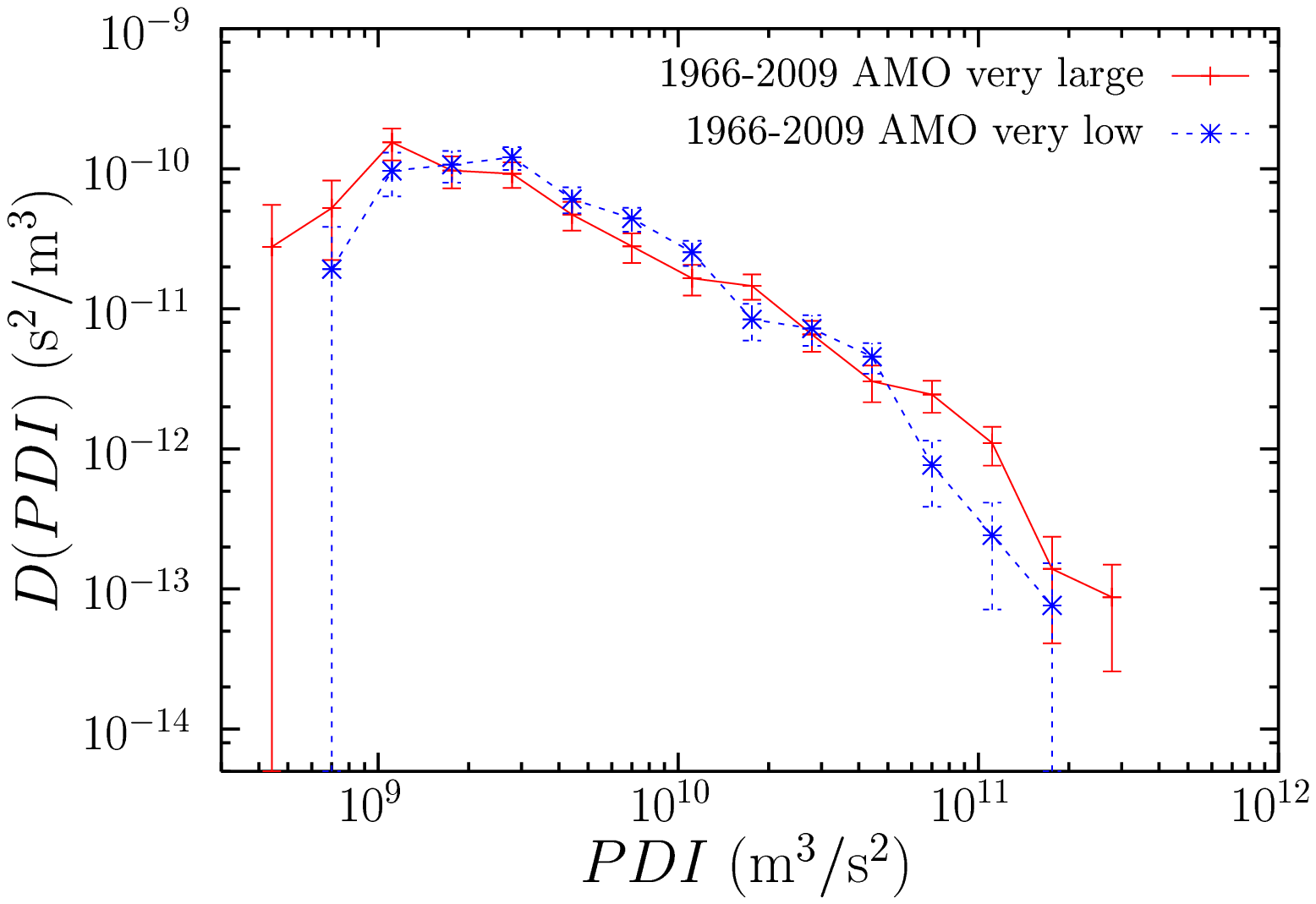}
(b) \\
 \noindent\includegraphics[width=20pc]{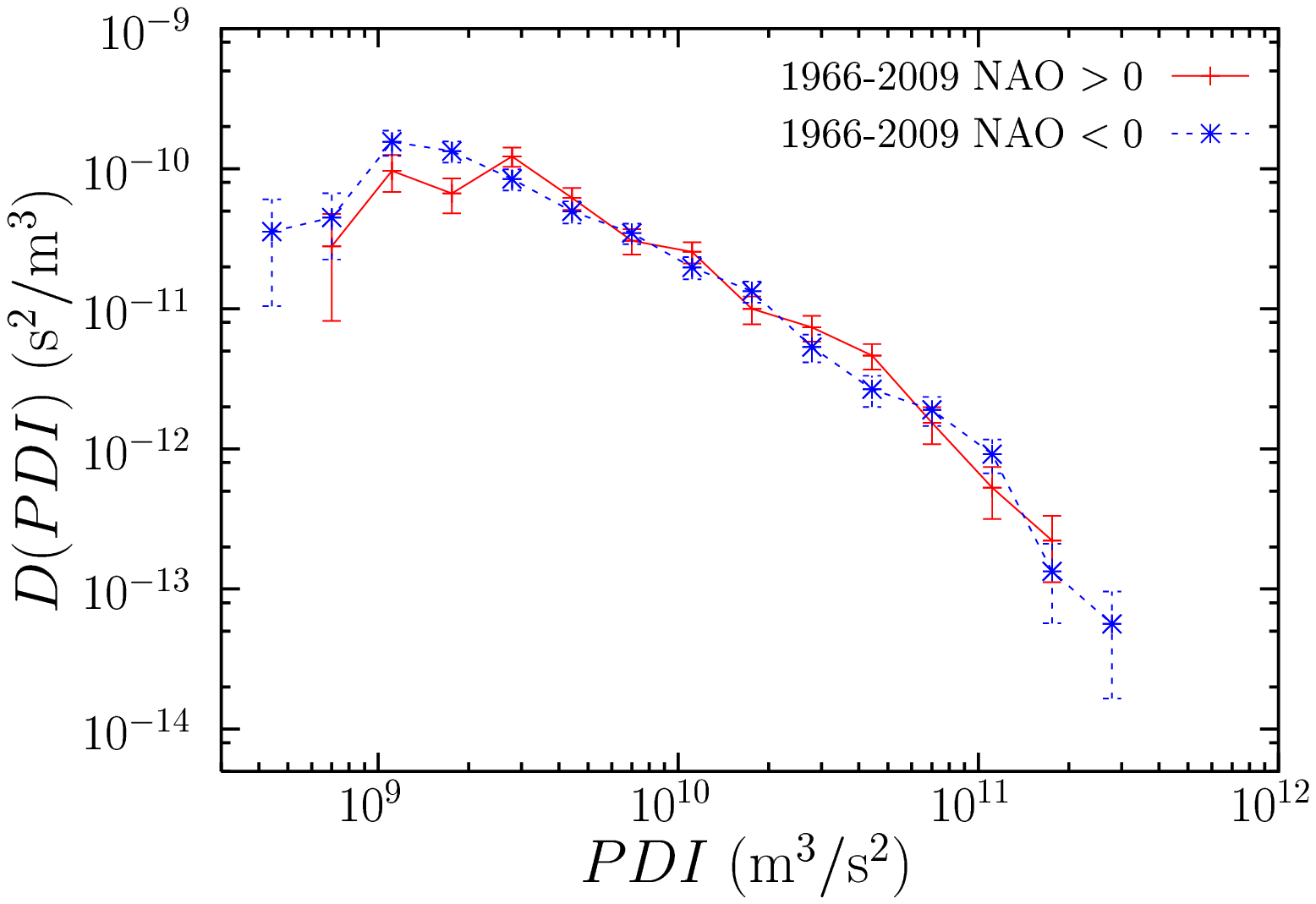}
(c) 
\caption{
(a) Same North-Atlantic PDI data separating this time by 
hurricane-season MEI 
above its mean plus $1/2$ of the standard deviation (labeled very large)
and below the mean minus $1/2$ of the standard deviation (very low).
(b) The same for AMO. 
(c) Same data separating by NAO $ > 0$ and NAO$ < 0$.
In this case the differences are clearly non significant.
\label{fig6}}
 \end{figure}

\begin{figure}
 \noindent\includegraphics[width=20pc]{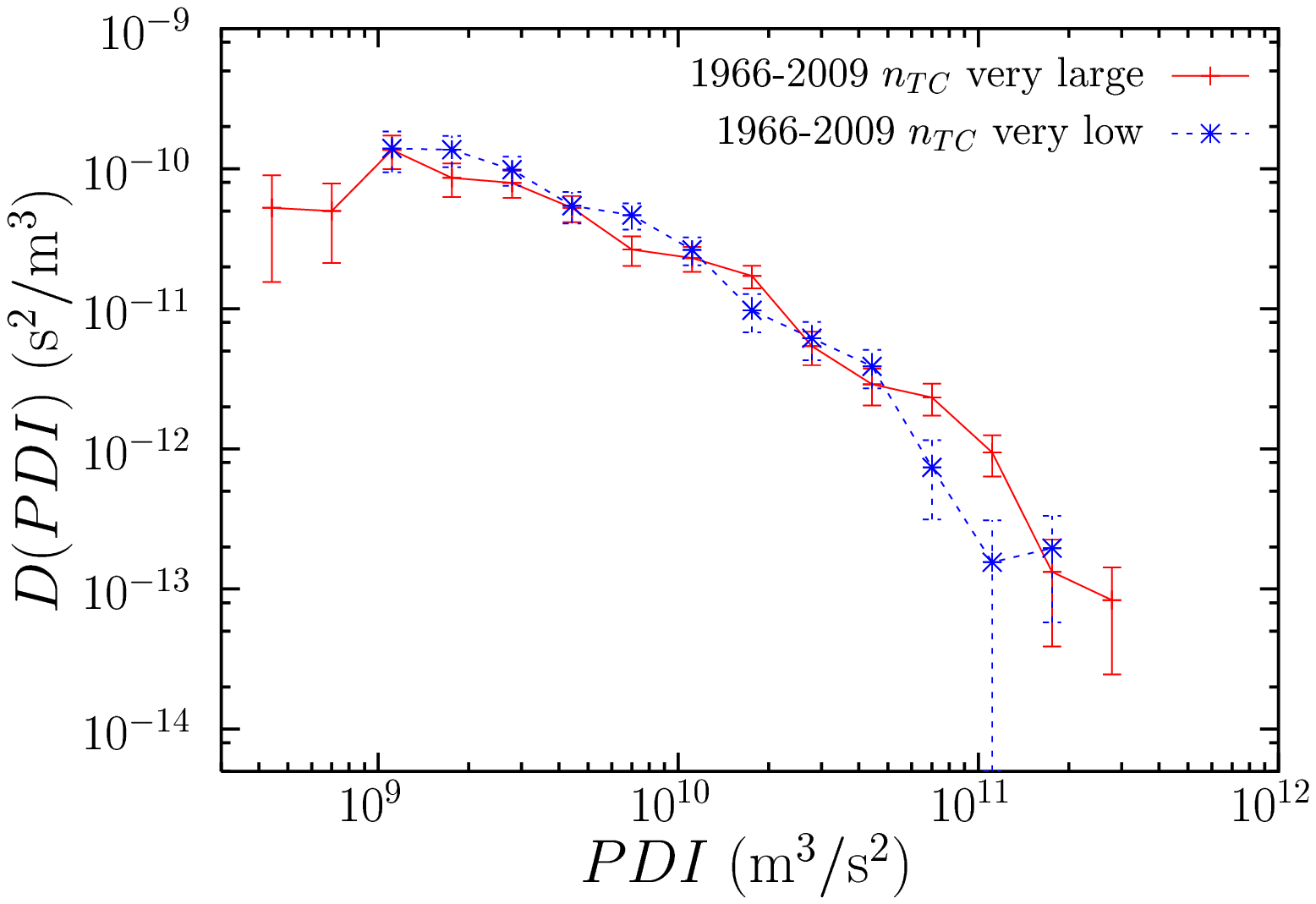}
(a) \\
\noindent\includegraphics[width=20pc]{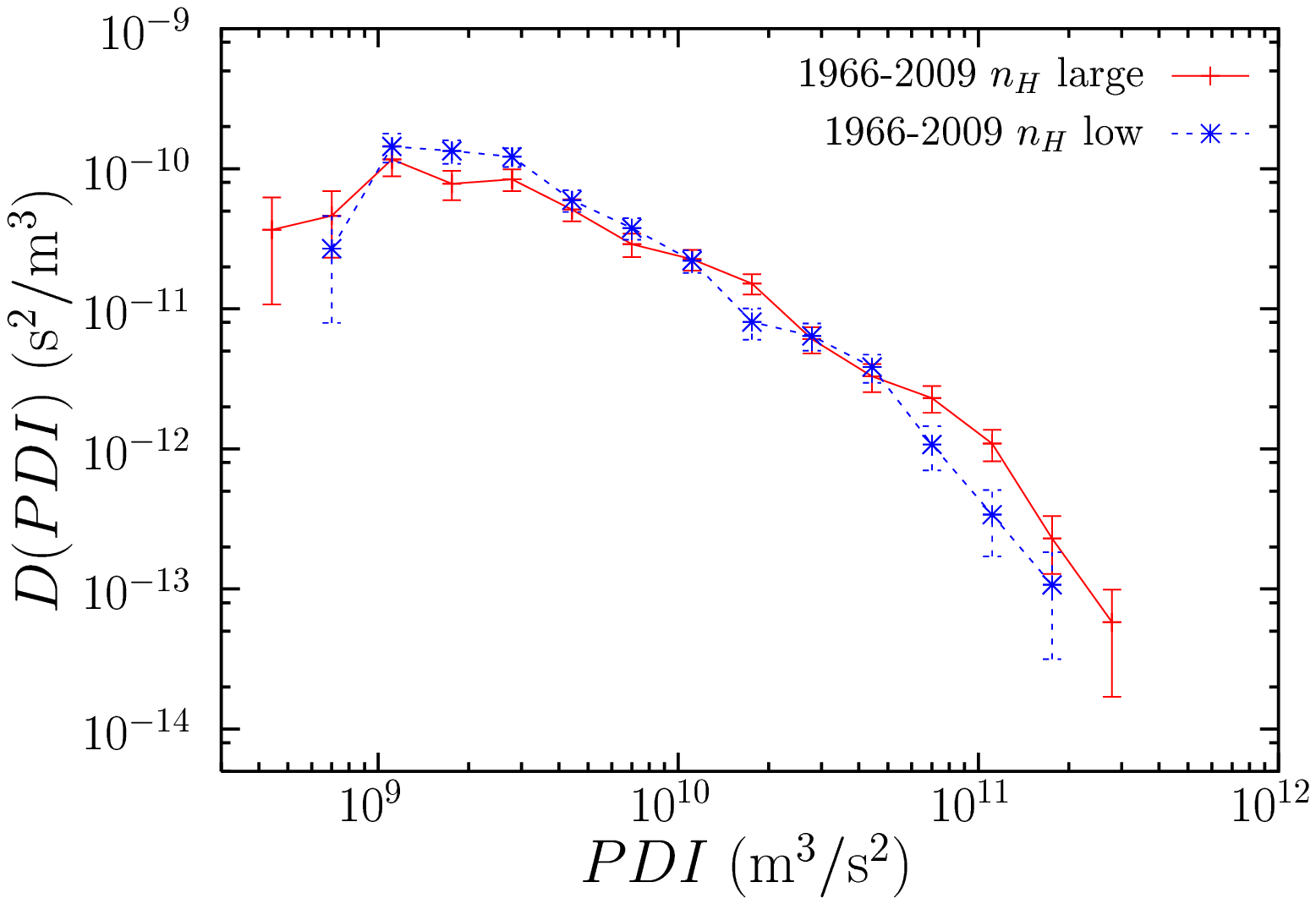}
(b)
 \caption{
(a) Same North-Atlantic data separating by annual number of
tropical cyclones above and below its mean $\pm$ $1/2$ of the standard deviation
(labeled as very large and very low, respectively).
(b) The same separating by number of hurricanes above or below its mean
(large and low).
This plot is not the one corresponding to the last column 
of table 3, as tropical storms are not removed here.
Caution has to be present in the interpretation of this case, see text.
\label{fig7}}
 \end{figure}

The results can be seen in Figs. \ref{fig6} and \ref{fig7} and in table 3.
Comparing PDI distributions for positive and negative values
of the ENSO, AMO, and NAO indices, 
only the ones separated by the AMO show
significant differences (at the 95 \% level),
with positive values of AMO triggering larger extreme 
PDI values.
The lack of influence of ENSO in the PDI is 
in agreement with \cite{Corral_hurricanes}.

But, if the PDI distributions are compared 
for high enough values of the indices, 
both the MEI and the AMO show a clear influence
on the PDIs (not the NAO).
As seen in Fig. \ref{fig6},
the presence of El Ni\~no (very high MEI), in addition to 
reduce the number of North Atlantic hurricanes
(which is well known),
decreases the value of their PDI.
Of course, 
just the same happens for very low values of the 
AMO index, which decrease both hurricane numbers\citep{Goldenberg_Science}
and the largest PDI values.
La Ni\~na or high values of the AMO have the opposite effect, 
increasing the most extreme PDIs.

The influence of the number of tropical cyclones in the values of the PDI
is not significant for values of $n_{TC}$ above or below its mean, 
but larger/smaller values of $n_{TC}$ (mean $\pm$ $1/2$ of standard deviation)
are correlated with larger/smaller values of the PDIs,
with an increase in $\langle PDI\rangle$ around 60 \%
when the mean $n_{TC}$ goes from 7.35 events to 16.8.
If we consider hurricane counts this effect is more pronounced, 
although care has to be taken with a kind of circular argument here:
just by chance, the number of hurricanes can increase
at the same time that the number of tropical storms decreases, 
then, it is likely that the mean PDI will be higher in this case
(as the maximum sustained wind speed is correlated with PDI), 
but no physical effect is behind this,
only statistical fluctuations.
Therefore, we study the influence of the number of hurricanes
on the PDI distributions of hurricanes only,
finding that years with more hurricanes
also have larger PDI values.
This seems to indicate that the conditions for genesis and survival of the storms
are correlated; compare with the conclusions in Sec. 4.3 of \cite{Elsner_Kara}.

%%% End of body of article:

%%%%%%%%%%%%%%%%%%%%%%%%%%%%%%%%
%% Optional Appendix goes here
%
%%%%%%%%%%%%%%%%%
% Geophysical Research Letters only allows an appendix without a letter.
%% You can get this result with
%  \section*{Appendix}
%  or
%  \section*{Appendix: Title}
%%%%%%%%%%%%%%%%%
%
% \appendix resets counters and redefines section heads
% but doesn't print anything.
% After typing  \appendix
%
% \section{Here Is Appendix Title}
% will print
% Appendix A: Here Is Appendix Title
%
% \section*{Appendix}
% will print
% Appendix
%
% \section*{Appendix: Here Is Appendix Title}
% will print
% Appendix: Here Is Appendix Title
%
% For only 1 appendix \appendix \section{Appendix} is preferred.
% which will print
% Appendix A

%%%%%%%%%%%%%%%%%%%%%%%%%%%%%%%%%%%%%%%%%%%%%%%%%%%%%%%%%%%%%%%%
%
% Optional Glossary or Notation section, goes here
%
%%%%%%%%%%%%%%
% Glossary only allowed in Reviews of Geophysics
% \section*{Glossary}
% \paragraph{Term}
% Term Definition here
%
%%%%%%%%%%%%%%
% Notation -- End each entry with a period.
% \begin{notation}
% Term & definition.\\
% Second Term & second definition.
% \end{notation}
%%%%%%%%%%%%%%%%%%%%%%%%%%%%%%%%%%%%%%%%%%%%%%%%%%%%%%%%%%%%%%%%
%
%  ACKNOWLEDGMENTS

\section{Summary and Conclusions}

%%%%%discutir la diferencia con nature physics!!!!!

We have illustrated how diverse metrics have been introduced
by several authors
in order to quantify tropical-cyclone activity,
for instance: major-hurricane counts, total major-hurricane days,
major-hurricane duration, NTC, etc...
In general, these metrics show a clear increase in activity in the North Atlantic since 1995
--although it is difficult to associate a single cause to this phenomenon.
Particularly interesting is the PDI, which was proposed by \cite{Emanuel_nature05}
in order to estimate annual tropical-cyclone kinetic energy dissipation
in different basins.

In a previous work \citep{Corral_hurricanes} 
we demonstrated how the use of the PDI to characterize the
energy dissipation of individual tropical cyclones gave coherent and robust results, 
leading to nearly-universal power-law distributions
(i.e., with roughly the same exponents for different ocean basins).
The outcome of this approach, in contrast to the findings obtained using other indices of 
activity, has important implications for the understanding of the
physics of tropical cyclones, as it allows a connection with 
self-organized-critical systems \citep{Corral_Elsner}.
This, in turn, opens interesting questions about the limits of 
predictability of tropical cyclones and, in a broader context, 
about the compatibility between criticality and chaos 
in the atmosphere.

Due to the important implications that the emergence of power-law distributions has, 
we also discuss different ways of fitting them.
The main problem is not to find the power-law exponent
(which is the only parameter of the distribution)
but to decide for which values of the variable the power-law 
regime holds. The procedure is based in the work of \cite{Clauset},
although some variations seem necessary in order to improve the performance
of the method.

Whereas the power-law part of the PDI distribution
does not change under different climatic conditions
(as it should be in a critical system),
the tail of the distribution does, containing therefore
precious information about external influences on the system, 
in particular on the finiteness of the part of the basin
able to sustain tropical-cyclone activity.
Therefore, a probability distribution modeling not only the power-law regime
but also the faster decay at the largest PDI values seems very appealing.
We propose the use of the gamma distribution, combining a power law
with an exponential decay.

In this way, it is shown that the cutoff parameter modeling the exponential decay
increases with annually averaged SST,
which means that high SST has an effect on the PDI that is analogous to expanding 
the effective size of the part of the ocean over which tropical cyclones develop
(and the opposite for low values of the annual SST).
The same effect is found for the AMO index, with high AMO values leading to 
larger extreme PDI values.
However the El Ni\~no phenomenon has the opposite effect, 
being the presence of La Ni\~na which triggers the largest (in dissipated energy)
North Atlantic hurricanes.
The number of hurricanes is also found to be positively correlated with their PDI.
In contrast, for the NAO index we do not find any significant correlation.

In conclusion, the characterization of PDI probability distribution reveals as an %%%%%% itself????
interesting tool not only to learn about the fundamental nature of tropical cyclones
but also to evaluate the effect of different climatic indices on the energy dissipated
by them.

\section*{Appendix: Relation between PDI and Dissipated Energy}

The power dissipated at any time by a tropical cyclone
can be estimated by means of the formula
\citep{Bister_Emanuel},
$$
P(t) = \int \rho C_D v^3 d^2r,
$$
where 
$t$ is time,
$r$ is the spatial coordinate
over the Earth's surface,
$\rho$ is the air density,
$C_D$ is the drag coefficient,
and $v$ is the modulus of the wind velocity
(the latter three depending on $r$ and $t$).

%In practice 
%it is necessary to simplify this formula.
The formula can be simplified in practice.
\cite{Emanuel_power} %%assumed
took representative constant values for the density
and the drag coefficient, 
$\rho= 1$ kg/m$^3$, $C_D=0.002$,
and assumed a simple velocity profile, 
$v(\vec r,t) =v_m(t) f(r/R(t)) $,
with $v_m(t)$ the maximum speed
across the storm at time $t$,
$R(t)$ some characteristic radius of the storm, 
as the radius of maximum winds, and $f$ 
a scaling function that is the same for all storms.
Denoting the spatial integral as $I=\int 2 \pi f^3(u) u du$,
then
$$
P(t)\simeq \rho C_D I  R^2(t) v_m^3(t) = k  R^2(t) v_m^3(t).
$$
%\cite{Emanuel_power} 
Emanuel does not provide enough details about the integral $I$,
but from the values of power, radius, and speed 
that he uses as an illustration, it has to be $I\simeq 2$, 
and then $k=0.004$ kg/m$^3$.

One can get an estimation of the dissipated kinetic energy $E$ just integrating $P(t)$ over time.
Introducing an averaged radius over the storm, $R_m$,
and the PDI discretization \citep{Emanuel_nature05},
$$
E = \int P(t) dt \simeq \rho C_D I R_m^2 \int v_m^3(t) dt= k  R_m^2 PDI.  
$$

The best-track records still do not provide the information required by
this equation, as in general the storm radius is missing.
However, for the Northwestern Pacific, since 2001, some radii are available
\citep{JTWC_data}. Out of 11358 6-hour records for the period 2001--2010, 
1671 provide non-zero values of the radius of maximum winds and the eye diameter.
Averaging these values of the radii of maximum winds we get $R_m\simeq 30$ km.
If we average the square of the radius of maximum winds and take the square root, 
then we get as a characteristic value $R_m \simeq 35$ km, 
which is the value we use for the North Atlantic, 
then, $k  R_m^2 = 4.9 \cdot 10^6 $ kg/m and
$$
E \simeq 4.9 \cdot 10^6 \, PDI,
$$
which yields dissipated energy in Joules if the PDI is in m$^3$/s$^2$.

Under these approximations, as the North-Atlantic individual-storm PDI ranges between
$5 \cdot 10^8$ and $2 \cdot 10^{11}$ m$^3$/s$^2$,
the estimated dissipated energy turns out to be roughly between
$3 \cdot 10^{15}$ and $10^{18}$ J,
i.e., between 0.6 and 200 megatons
(1 megaton $=4.18 \cdot 10^{15}$ J).
Nevertheless, the real range of variation will be larger, 
as the variability of the radius, which we have disregarded, 
increases the variability of the dissipated energy.

\begin{acknowledgments}
The authors acknowledge initial support by J. E. Llebot
and guidance by J. Kossin. First results of this
research came through collaboration with A. Oss\'o.
I. Blad\'e,
A. Clauset, A. Deluca, K. Emanuel, R. D. Malmgren, and N. Moloney %%%, and A. Turiel
provided some feedback on different parts of the research.
Money and a one-year bursary for A. Oss\'o was obtained from
Explora-Ingenio 2010 program (MICINN, Spain), grant FIS2007-29088-E.
Other grants are
FIS2009-09508 and 2009SGR-164. The first author is also a participant of the
Consolider i-Math project.
\end{acknowledgments}

\end{article}

%% Enter Figures and Tables here:

% When submitting articles through the GEMS system:
% COMMENT OUT ANY COMMANDS THAT INCLUDE GRAPHICS.

% Figure captions go below this illustration; Table captions go above tables

% ONE-COLUMN figure/table, including eps graphics
%
% \begin{figure}
% \noindent\includegraphics[width=20pc]{samplefigure.eps}
% \caption{Caption text here}
% \end{figure}
% \end{document}
%
% \begin{table}
% \caption{}
% \end{table}
%
% ---------------
% TWO-COLUMN figure/table
%
% \begin{figure*}
% \noindent\includegraphics[width=39pc]{samplefigure.eps}
% \caption{Caption text here}
% \end{figure*}
%
% \begin{table*}
% \caption{Caption text here}
% \end{table*}
%
% see below for how to make landscape figures or tables

%%% End the article here:

\end{document}